\documentclass{aa}

\usepackage{natbib,epsfig,amsmath,amssymb}

\def\ga{\,\hbox{\hbox{$ > $}\kern -0.8em \lower 1.0ex\hbox{$\sim$}}\,}
\def\la{\,\hbox{\hbox{$ < $}\kern -0.8em \lower 1.0ex\hbox{$\sim$}}\,}
\def\beq{\begin{equation}}
\def\eeq{\end{equation}}

\titlerunning{Spiral-driven accretion in protoplanetary discs}
\authorrunning{Hennebelle et al.}

\begin{document}

\title{Spiral-driven accretion in protoplanetary discs - III
tri-dimensional simulations}

\author{Patrick Hennebelle\inst{1,2}, Geoffroy Lesur \inst{3,4}, S\'ebastien Fromang \inst{1}}
\institute{
Laboratoire AIM, 
Paris-Saclay, CEA/IRFU/SAp - CNRS - Universit\'e Paris Diderot, 91191, 
Gif-sur-Yvette Cedex, France \\
\and
LERMA (UMR CNRS 8112), Ecole Normale Sup\'erieure, 75231 Paris Cedex, France
\and Univ. Grenoble Alpes, IPAG, 38000, Grenoble, France
\and CNRS, IPAG, F-38000 Grenoble, France}

\abstract{Understanding how accretion proceeds in proto-planetary discs and more generally their dynamics is a crucial 
issue for explaining the conditions in which planets form.} {The role that accretion of gas from the surrounding molecular cloud  onto the disc may have 
on its structure needs to be  quantified.} {We perform tri-dimensional simulations using the Cartesian AMR code RAMSES of an 
accretion disc subject to infalling material.} 
{For the aspect ratio of $H/R \simeq 0.15$ and disk mass $M_d \simeq 10^{-2}$ M$_\odot$ used in our study, we find 
that for typical accretion rates on the order of a few 10$^{-7}$ M$_\odot$ yr$^{-1}$, values of the $\alpha$ parameter as high as a
few 10$^{-3}$ are inferred. The mass that is accreted in the inner part of the disc is typically  at least $50\%$ 
of the total mass that has been accreted 
onto the disc.} {Our results suggest that external accretion of gas at moderate values, 
onto circumstellar discs may trigger prominent spiral arms, reminiscent of recent observations made with various instruments,
 and lead to significant transport through the disc. If confirmed from observational studies, 
such accretion may therefore influence disc evolution.} 

\maketitle

\section{Introduction}

Accretion discs are ubiquitous  in astrophysics as they are
observed around stars and black holes and as such they have received 
considerable attention. In particular 
it is widely admitted that local instabilities 
such as the magneto-rotational instability \citep[MRI, e.g.][]{balbus2003} or  
the gravitational instability \citep[e.g.][]{lodatorice2004} are 
responsible within different contexts 
for triggering the transport of angular momentum and mass. 
The conditions under which these mechanisms may actually work remain
debated, for example  it is not clear that protostellar 
discs are sufficiently ionized for the MRI to operate efficiently \citep[e.g.][]{lesur+2014,turner2014}.

 Mass transferring binary systems are a well studied type of disc and it has early been recognized 
that the spiral shocks induced by the presence of the companion and the accretion
 is a source of external disturbances, which can 
trigger significant accretion within the disc 
  \citep[][]{spruit87, larson1990, vishniac1989}. This has been confirmed by  
 a series of numerical simulations
\citep{sawada1987,spruit1987b,rozyczka1993,yukawa1997}. In their recent study,
\citet{ju+2016} considered the impact of spiral shocks in the presence of 
MRI turbulence and concluded that the former are playing a significant role
also in this context.

On the contrary, most studies have been studying proto-planetary discs in isolation. 
This is because it is not clear yet whether these objects are still subject 
to infall from their surrounding parent molecular cloud.
 A notable exception regards
the studies which have addressed the question of disc formation out of their parent dense cores. 
 The newly formed discs  are usually massive, self-gravitating and often 
strongly magnetized
and it is generally admitted that  angular momentum can  be transported 
by the gravitational torque or by magnetic braking \citep[e.g.][]{vorobyov2008,machida+2010,joos+2012,li+2013,vorobyov2015,masson+2016}. 
For example \citet{vorobyov2015} performed a series of bidimensional studies of
self-gravitating and viscous discs embedded into their parent cores and show that the 
infall of material has a drastic impact on their evolution. In particular, 
the angular momentum of the infalling material, appears to be particularly
important  on the evolution.


To investigate specifically the role that infall may have on the disc, \citet{lesur+2015} 
performed bidimensional hydrodynamical simulations. They found that infall 
induces high values of $\alpha$ in the outer part of the disc and smaller but non 
negligeable values, on the order of a few 10$^{-4}$ for an imposed accretion rate of about 10$^{-7}$ M$_\odot$ yr$^{-1}$
 in the inner part of the disc where  a plateau of constant $\alpha$ seems to be reached.
It has also been shown that although slightly less efficient than non axisymmetric inflows, 
axisymmetric flows, quickly break up   forming a spiral pattern and  trigger even  smaller
 scale unstationary structures within the disc.

It is now necessary to investigate this issue in three dimensions, and for that purpose
 we present here tridimensional numerical simulations of a disc subject 
to infall.
Magnetic field  and self-gravity are explicitly neglected at this stage. 
The second section of the paper describes the numerical setup, 
including initial and boundary conditions. It also presents
a simulation of a non-accreting disc which allows us to assess the initial conditions 
and the effect of numerical viscosity.
In the third section, we present in detail a simulation and 
we study the influence of numerical resolution. We provide some qualitative comparisons
with previous analytical works.
The  influence of the 
strength of the accretion and quantitative comparison with bidimensional simulations 
are provided in the fourth section.
The influence of the geometry of the accretion flows as well as of the amount of angular momentum
it carries is presented in the fifth section. 
Section six provides some discussions making the link with recent observations of spiral patterns
in transitional discs and their interpretation. 
The seventh section concludes the paper.

\section{Tridimensional simulations: setup, resolution and runs performed}
We now turn to the description of the series of 3-dimensional 
simulations that we performed. In a companion paper \citep{lesur+2015}, we
presented a series of 2-dimensional runs.

\setlength{\unitlength}{1cm}
\begin{figure} 
\begin{picture} (0,8)
\put(0,0){\includegraphics[width=8cm]{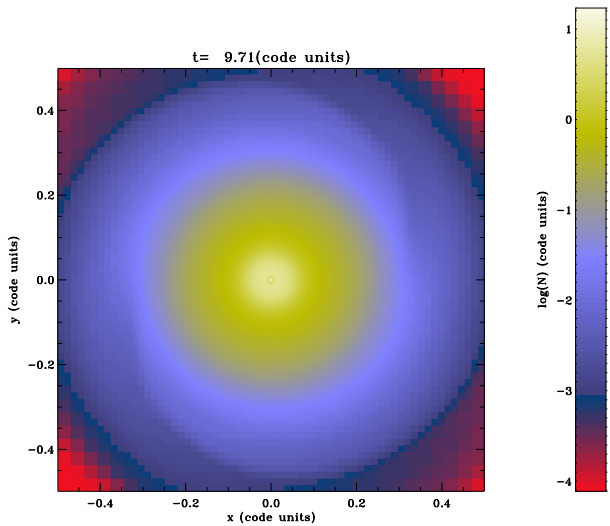}}  
\end{picture}
\caption{No accretion case. Column density map. The disc has a radius of about 
$r= 0.25$. }
\label{no_accret_map}
\end{figure}

\setlength{\unitlength}{1cm}
\begin{figure} 
\begin{picture} (0,9)
\put(0,4.5){\includegraphics[width=9cm]{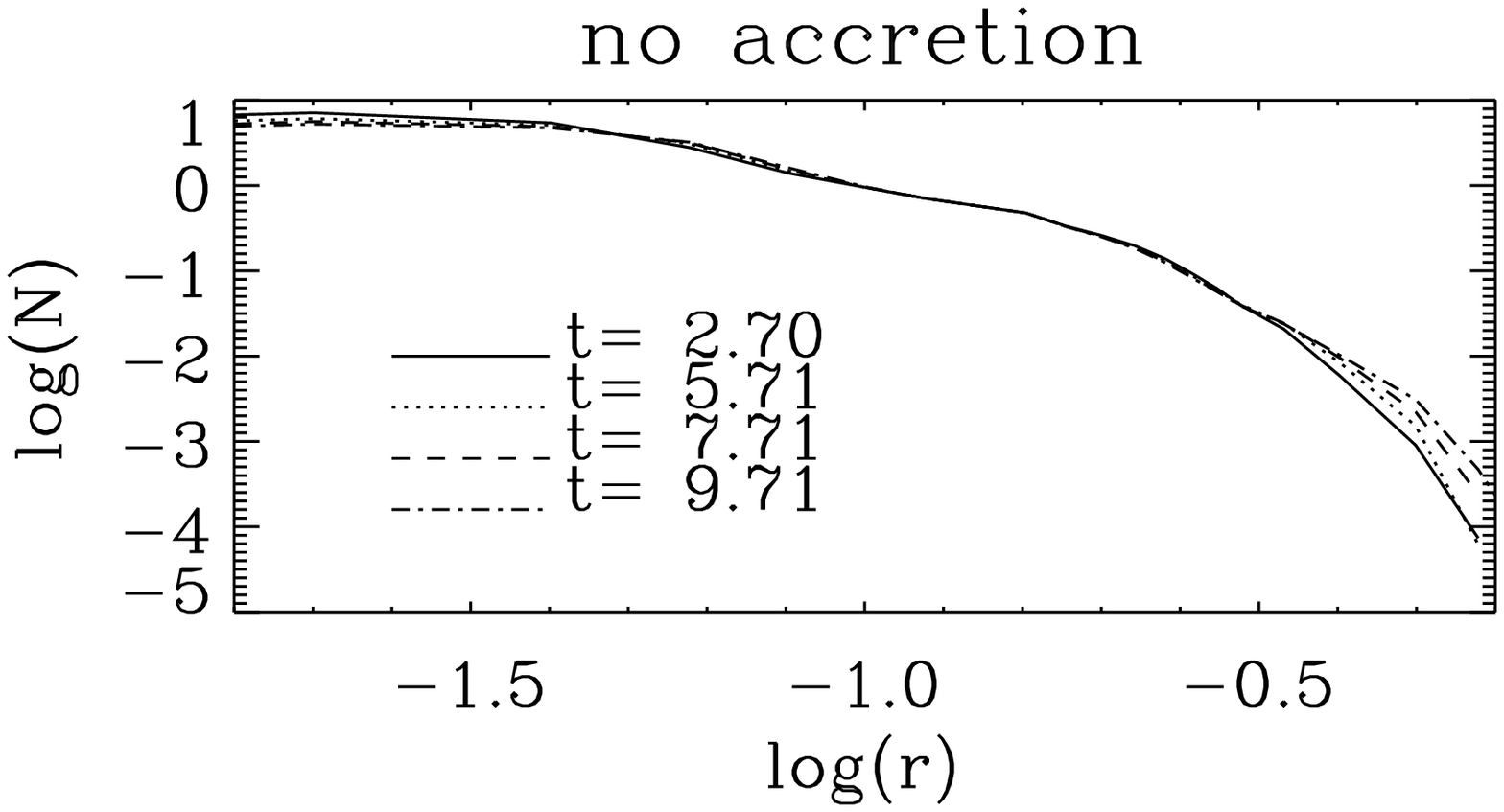}}  
\put(0,0){\includegraphics[width=9cm]{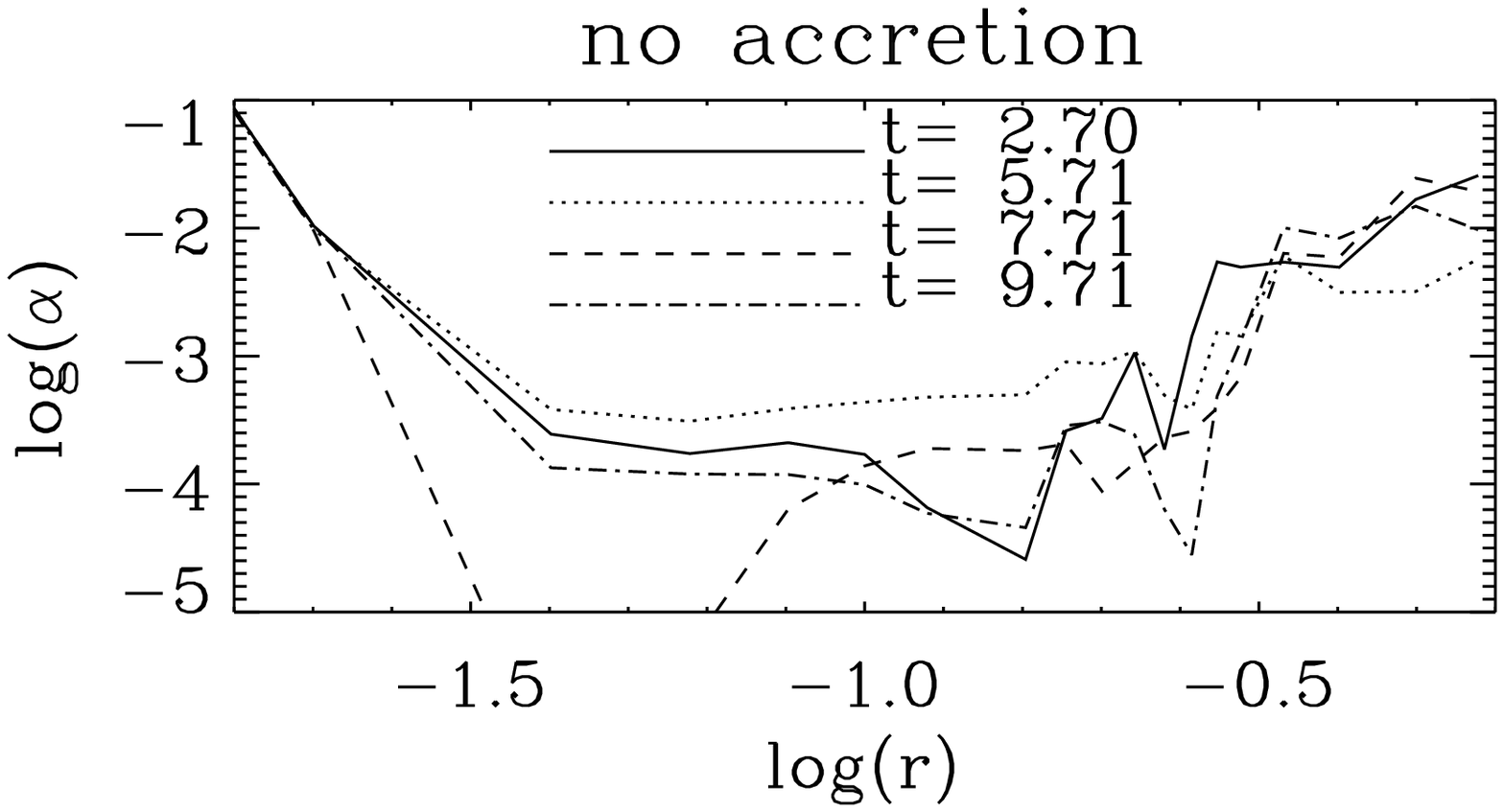}}  
\end{picture}
\caption{No accretion case. Column density  (top panel) and $\alpha$ value (bottom panel)
as a function of the radius $r$.
Typical variations remains limited and the $\alpha$ values are below $10^{-3}$ 
rapidly falling below $3 \times 10^{-4}$. Note that at small radii ($log(r) < -1.6$), $\alpha$ increases to high values because of low resolution.}
\label{no_accret}
\end{figure}

\subsection{General setup}

\subsubsection{Code}
To perform our simulations, we employ the code RAMSES
\citep{teyssier2002}, which is an adaptive mesh refinement 
code working in Cartesian geometry and using finite 
volume methods and Godunov solvers.
Cylindrical or spherical geometries have been traditionally used 
in the context of disc studies because they ensure exact 
conservation of angular momentum. This is also 
the choice which has been made for the simulations presented 
in \citet{lesur+2015}.
However, these geometries also introduce singularities along the z-axis or
at the origin, that requires specific treatments, while 
Cartesian codes do not present such a difficulty.

\subsubsection{Initial conditions}
 The initial state is an axisymmetric equilibrium between gravitational, centrifugal and pressure forces.
We adopt the locally isothermal approximation, 
\begin{eqnarray}
\label{sound_speed}
C_s = C_s^0 \times \left( { r \over r_d} \right)^{-1/2},
\end{eqnarray}
where $r$ is the cylindrical radius and $C_s$ is the sound speed with  $C_s^0= \sqrt{0.1}$.
 This approximation is supported by detailed models of discs \citep{dalessio1998,armitage2011} temperature 
structure which in particular show that the dominant source of heating above 2 AU is due to the star irradiation.
The mass of the central star is $M_*=1$ and the disc radius is equal to $r_d=1/4=L_b / 8$
where $L_b=2$ is the size of the computational box. 
This choice ensures that the size of the disc is sufficiently small with respect to the computational box 
size, in such a way that it is not significantly influenced by boundary conditions. 
Below only the central part of the computational box is displayed. Therefore the disc occupies initially half of the  image displayed. 
In these units, $G$, the gravitational constant is equal to 1.
This ensures that in the radial direction the gravitational force is always 
10 times larger than the pressure one. The disc aspect ratio is given by 
\begin{eqnarray}
{h \over r} \simeq {C_s \over r \Omega} =  { C_s^0 \times \left(  r_d \right)^{1/2} \over \sqrt{G M_*} } \simeq 0.15,
\end{eqnarray}
where we used  $\Omega \simeq r^{-3/2}$ (since $M_*=1$ and $G=1$).
With this choice, the aspect ratio is therefore constant and equal to about 0.15.

Solving for the  axial mechanical equilibrium, we obtain the density 
profile:
\begin{eqnarray}
\label{rho_ini}
\rho = \rho_c(r) \exp \left( {G M_* \over C_s^2}  \left( {1 \over \sqrt{r^2+z^2}} - { 1 \over r}\right)\right),
\end{eqnarray}
where $\rho_c$ is a free function of $r$ that we have chosen to be equal to
\begin{eqnarray}
\label{rho_central}
\rho_c(r) = { \rho _c^0 \over (r/r_d)^{2.5} }.  
\end{eqnarray}

The rotation curve is then simply obtained by requiring mechanical equilibrium 
in the radial direction
\begin{eqnarray}
\label{rotat}
\Omega ^2 = {\partial_r (C_s^2 \rho) \over r \rho} + {G M_* \over (r^2+z^2)^{3/2}}
\end{eqnarray}

The equilibrium solution is truncated at $r_d$ where the density is decreased to a value 
one hundred times lower. This causes the disc to expand initially. However, as shown 
below the effect remains limited even in the absence of any external accretion.

The orbital time at the initial disc external radius, $r_d$, is 
simply given by
\begin{eqnarray}
\label{time_scale}
\tau _d = {2 \pi \over \Omega} =  {2 \pi r_d^{3/2} \over G M_* } = {\pi \over 4} \simeq 0.78.
\end{eqnarray}
This will constitute the natural time units for the simulation results to which we will refer. 

\subsubsection{Inner boundary conditions}
The central star is placed at the center of the grid and the corresponding 
gravitational force is directly added to the gas smoothing the $r^{-2}$
dependence within a few computing cells.
There is no sink of matter associated and therefore the gas is simply pilling up at the vicinity of the star 
as the simulation evolves. Around the central star,  mesh effects are very important and 
the flow structure is much affected by the numerical resolution. As quantitatively discussed  later, 
this is only at a certain distance 
from the central object that the results can be considered as meaningful. 

The sound speed dependence, as stated by Eq.~(\ref{sound_speed}), leads to a singular behaviour 
near the z-axis. This creates spurious behaviours and leads to large velocities, which 
spoil the disc equilibrium. Therefore we have modified the sound speed 
at small radii, $r$. We essentially impose that below a radius $r_d/10$, the temperature 
is simply constant and equal to the value it has at this radius.  Since as described below, 
numerical resolution drops at high altitude, we also make the radius inside which the temperature 
remains constant, slowly increasing with height, giving it a parabolic shape.

\subsubsection{Outer boundary conditions}
Since  Cartesian coordinates are employed, the computational domain is 
a cubic box which causes important boundary effects. This would typically 
cause strong reflections and spurious $m=4$ modes that  affect the disc structure. 
To avoid these artefacts, the size of the disc has been chosen to represent only 1/4 of the 
whole box size as mentioned above, which is possible thanks to the AMR scheme of RAMSES. 
However, although it helps to reduce these effect, we find that they are still significant.
Therefore we impose at every timestep a uniform density and a Keplerian 
velocity for every cell located at $r>0.9$ (since the box size is $L_b=2$, the boundary cells 
in the equatorial plan are located between $r=1$ and $\sqrt{2}$) from the box center.
With this boundary condition, we found that in the absence of infall the flow remains sufficiently symmetrical 
as shown by Fig.~\ref{no_accret_map}.

\subsubsection{Numerical resolution}
\label{num_reso}
To perform our calculations efficiently we took advantage of the AMR scheme by 
using a coarse resolution outside the disc, that is to say at radius $r > r_d = 1/4$
and height larger than $z> r_d/2$, while using a finer resolution inside the disc. 
For most of the runs, the resolution employed the AMR level, $l=10$ inside the disc
implying that the disc radius is described by 128 cells and the whole disc 
lays in a box of $256 \times 256 \times 128 $ cells. 
As discussed above, the aspect ratio of the disc is about 0.15. Therefore at the disc
radius, the disc scale height is described by about 40 cells. At a radius of 
$r=r_d/10$, this number becomes equal to only 4 cells 
in a typical scale height. These simulations therefore can tackle a range of radius
on the order of $\simeq$5-10. Later we refer to this resolution as "standard''.

To investigate the issue of numerical resolution, we have also performed 
 runs with a higher resolution in the inner part. The area 
located at $r < 0.5 \times r_d $ is described using level $l=11$ implying that at 
$r=r_d/10$, the scale height is described by about 8 cells. We refer to this
resolution as ``high''. Note that at the same time, the radius at which the 
gas becomes isothermal has also been reduced by a factor 2.

For all simulations, above the disc radius, $r_d$ the 
resolution drops to level $l=8$. Then at radius $r >1.4 r_d = 0.35$ it drops to 
level $l=7$ and then at $r > 2 r_d = 0.5$, to $l=6$.

Typically the simulations request about 
 10,000 
(for the ``standard'' resolution) to about 
 100,000 cpu hours  (for ``high'' resolution).

\subsubsection{The accretion flow}
To describe the external accretion, we impose two types of inflow boundary conditions, 
one axisymmetric and one asymmetrical. 

For the asymmetrical accretion scheme, the gas is injected from a 
circle of radius $r_d$, contained in the $(y,z)$ plane, and centered in $x=-3.5 r_d, \, y=0, \, z=0$ 
at a radial velocity $U_{acc}$ and with a density $\rho _{acc}$. 
The accreted gas possesses a specific angular momentum that is equal
to either 0 or to $r_d ^2 \Omega(r_d)$, that is to say the angular momentum 
which corresponds to the keplerian velocity at the disc radius, $r_d$.
The accretion velocity is expressed as a function of the sound 
speed at the boundary, which is equal to $C_s=C_s^0/2$.
The accretion rate is thus given by
\begin{eqnarray}
\label{accret}
\dot{M} = \pi r_d ^2 U_0 {C_s^0 \over 2} \rho _{acc}.
\end{eqnarray}

For the axisymmetrical accretion scheme, the gas is injected from a ring 
of radius 3.5$r_d$ and half thickness $r_d$.
While the value of $\rho _{acc}$ is unchanged, the values of 
$U_{acc}$ are adjusted in such a way that the total accretion rate is 
identical to the asymmetrical case
\begin{eqnarray}
\label{accret_sym}
\dot{M} = 2 \pi (3.5 r_ d) \times (2 r_d) { U_0 \over 2 \times 2 \times 3.5} {C_s^0 \over 2} \rho _{acc}.
\end{eqnarray}
Thus for  given values of $\rho _{acc}$ and $U_0$, the value of $\dot{M}$ is identical for the 
two geometries.

Below we consider values of $U_0=0.5$ and 2. 
To determine typical values of $\rho _{acc}$, we must refer to observational case. 
We believe that an efficient way to characterize external infall in  accretion discs
is given by the dimensionaless number, $P$,
\begin{eqnarray}
\label{P_adim}
P = { 2 \pi \dot{M} \over M_d \Omega(r_d)} = { \dot{M} \tau_d \over M_d},
\end{eqnarray}
 that is to say the ratio between the mass accreted (within the computing box) over 
one orbital period at $r=r_d$ and the disk mass. This parameter describes the rapidity 
at which mass is delivered into the disc per disc mass and typical disc timescale.

For a typical T-tauri star of 1 $M_\odot$ and a radius $r_d=100$ AU,  
accretion rates onto the central star, $\dot{M}_{acc}$, on the order of up to  $10^{-7}$ $M_\odot$ yr$^{-1}$
have been inferred \citep{muzerolle2005}.  If we assume that accretion onto the star 
is a consequence of  external infall onto the disc, this suggests that the 
two accretion rates should be related maybe  comparable.
In the case of our simulations, with these numbers we have 
\begin{eqnarray}
\label{P_ttauri}
P &=& { 2 \pi \dot{M} \over M_d \Omega(r_d)} 
\simeq  10^{-2} { \dot{M} \over 10^{-7} M_\odot \, yr^{-1} } \\
\nonumber
&\times& \left( { M_d \over 10^{-2} M_\odot} \right)^{-1}
 \left( { M_* \over 1 \, M_\odot}  \right)^{-1/2}  \left( { r_d \over 100 \, AU}  \right)^{3/2}.
\end{eqnarray}

With the density profile stated by Eq.~(\ref{rho_ini}) and $\rho _c^0=1$, the 
mass of the disc initially is about 0.1. Since $r_d=1/4$, we find 
\begin{eqnarray}
\label{P_code}
P = { 2 \pi \dot{M} \over M_d \Omega(r_d)}  = { \pi^2 C_s^0  r_d^{7/2} \over M_d} U_0 \rho _{acc}
 \simeq  2 \times 10^{-2}  \, U_0 {\rho_{acc} \over 0.1 }. 
\end{eqnarray}

In the following, we will adopt  $\rho _ {acc}=0.1$. 
We typically integrate during about 15 orbits at the outer radius. 
At $r_d/10$, it roughly corresponds to about 500 orbits. 
During this amount of time and for the value $U_0=2$, the total mass inside the computational 
box  increases by about 30$\%$. We stress however, that the increase of the disc mass, that 
is to say inside the radius, $r_d$, is 3-4 times lower than this value. This means that the effective 
mass flux, which falls into the disc is 3-4 times lower than the analytical value stated by 
Eq.~(\ref{accret}). This is because most of the mass added in the computational box 
piles up in the external medium where the density is low initially.

\subsection{Runs performed}

Table~\ref{run} summarizes the runs performed in the paper
and provides consistent labels.
\begin{table}
\begin{tabular} {| {l} | {l} | {l} | {l} | {l} | {l} |  {l} | }
\hline
 label   & accret & $U_0$ & $\rho _{acc}$ & $P$ & $V_0$ & resolution \\
\hline
 NA   & no  & 0 & 0 & 0 & -  & standard \\
\hline 
 ASR2   & a  & 2 & 0.1 & 0.04 & 1  & standard \\
\hline 
ASR2h   & a  & 2 & 0.1 & 0.04 & 1  & high \\
\hline 
ASR0.5h   & a  & 0.5 & 0.1 & 0.01 & 1  & high \\
\hline  
SR2   & s  & 2 & 0.1  & 0.04 & 1  & standard \\
\hline 
ASNR   & a  & 2 & 0.1  & 0.04 & 0  & standard \\
\hline 
\end{tabular}
\caption{Summary of the different runs performed in the paper.
accret 'a' means asymmetrical accretion while 's' means symmetrical.
$U_0$ is the value of the accretion speed and 
$\rho _{acc}$ the accretion density (see Eq.~\ref{accret}).
$P$ is the accretion parameter (see Eq.~\ref{P_adim}).
$V_0$ is the rotation velocity of the accretion flow.
The standard and high  resolution are 
described in the text.}
\label{run}
\end{table}
This set of simulations attempts to explore the influence of parameters
regarding the accretion flow (velocity, density, symmetry and angular momentum), 
and the numerical resolution. 

As can be seen the smallest value of $P$ is $10^{-2}$, which for a 
$10^{-2}$ M$_ \odot$ disc and a 1 M$_ \odot$ star corresponds to an accretion 
of about 10$^{-7}$ M$_\odot$ s$^{-1}$. This corresponds to somewhat large 
values of accretion rate but as explained below, the effective accretion rate onto 
the disc is typically 3-4 smaller than this value. 
Note that reducing further the accretion rate would be difficult as 
numerical resolution (see Fig.~\ref{no_accret})
is putting constraints on the smallest flux that can be modelled.

\setlength{\unitlength}{1cm}
\begin{figure*} 
\begin{picture} (0,21)
\put(8,14){\includegraphics[width=8cm]{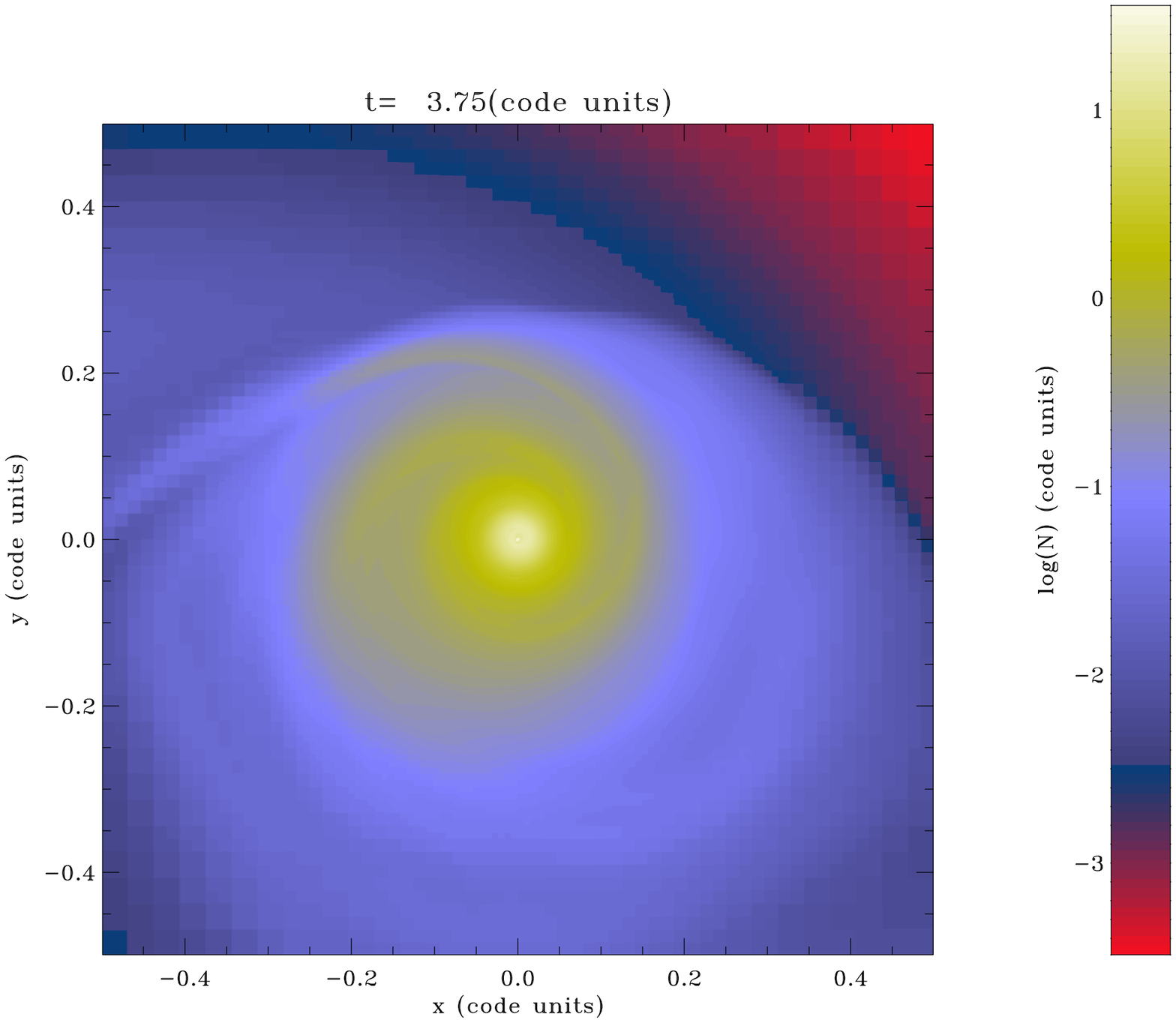}}  
\put(8,7){\includegraphics[width=8cm]{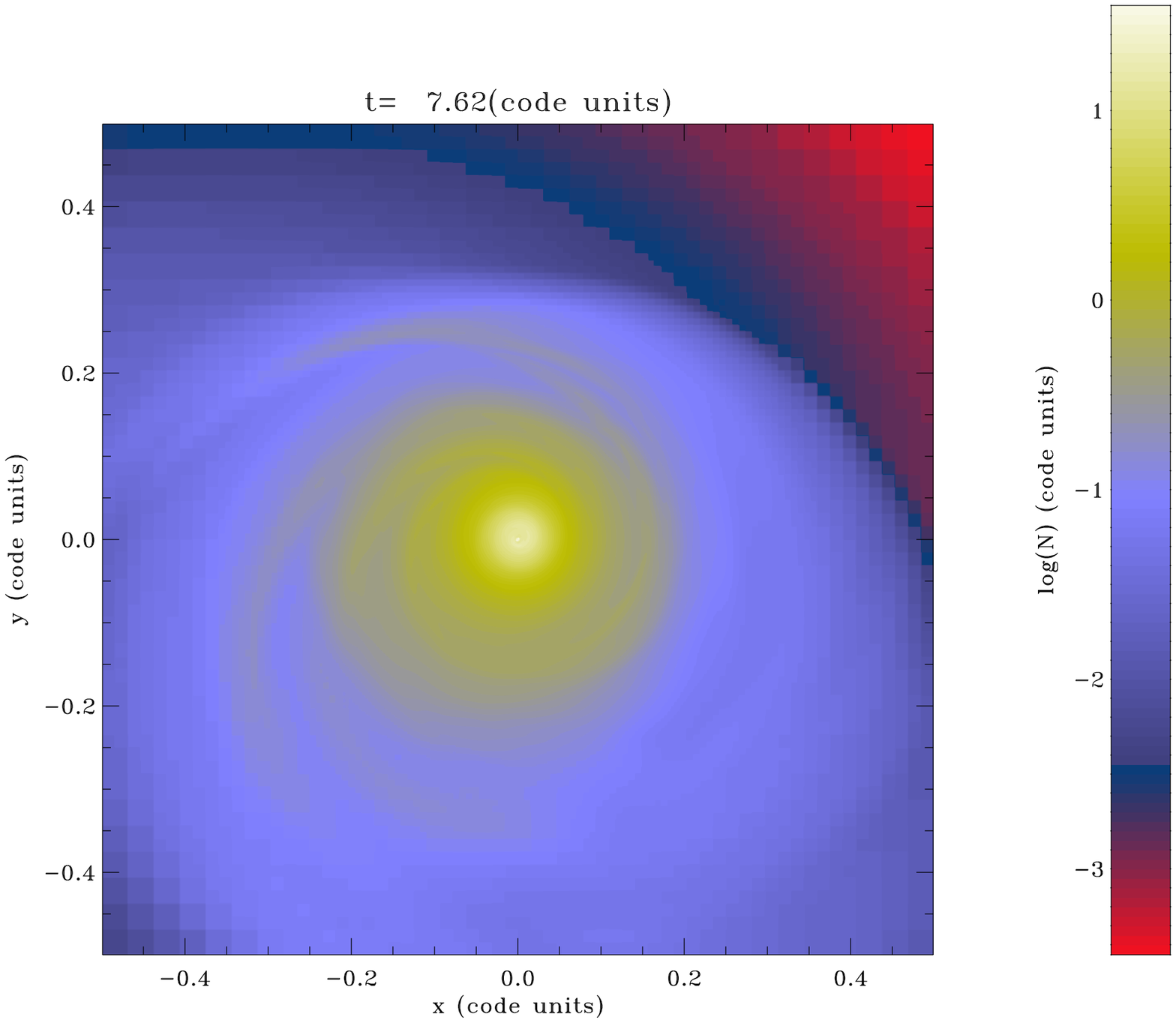}}  
\put(8,0){\includegraphics[width=8cm]{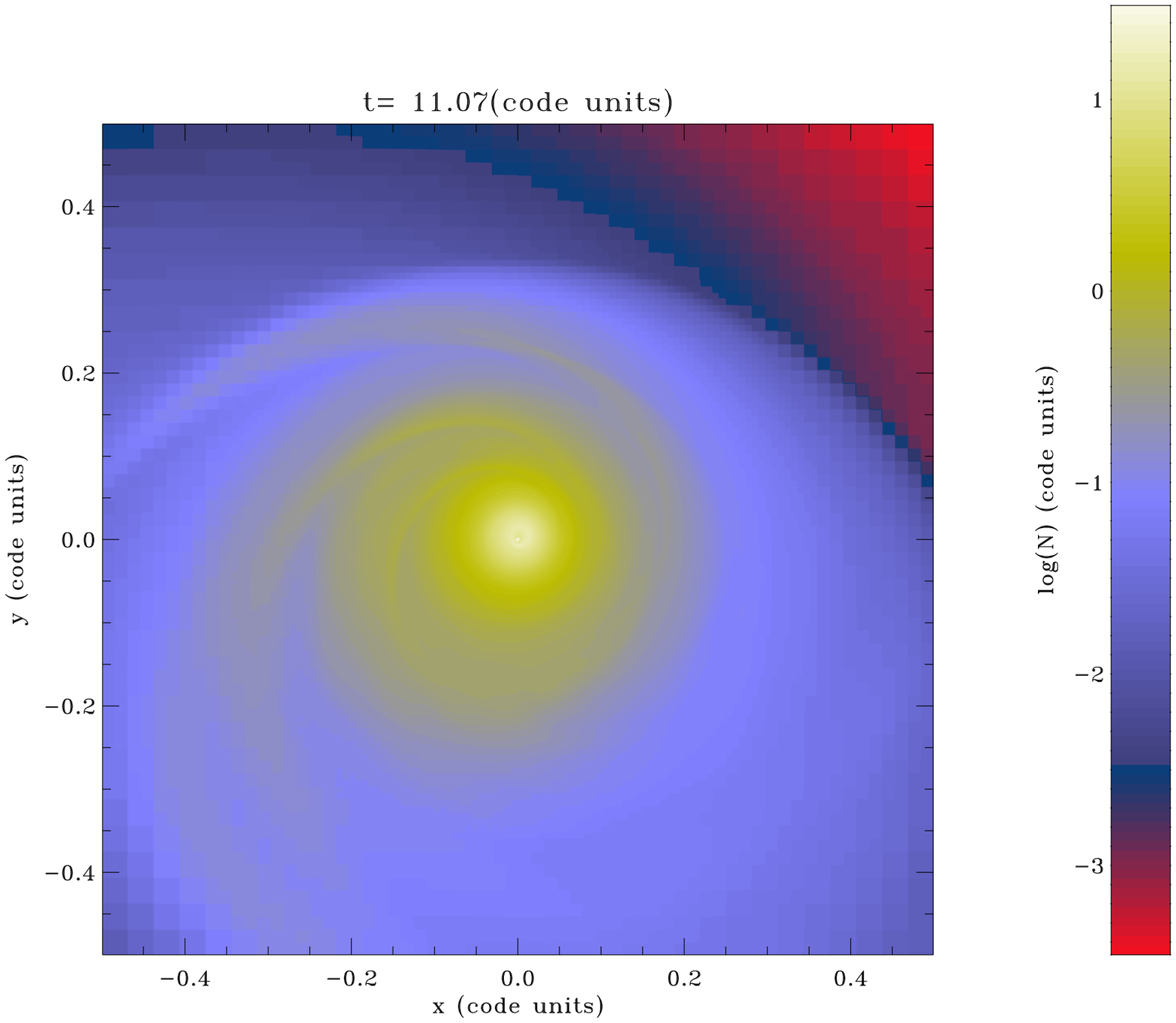}}  
\put(0,14){\includegraphics[width=8cm]{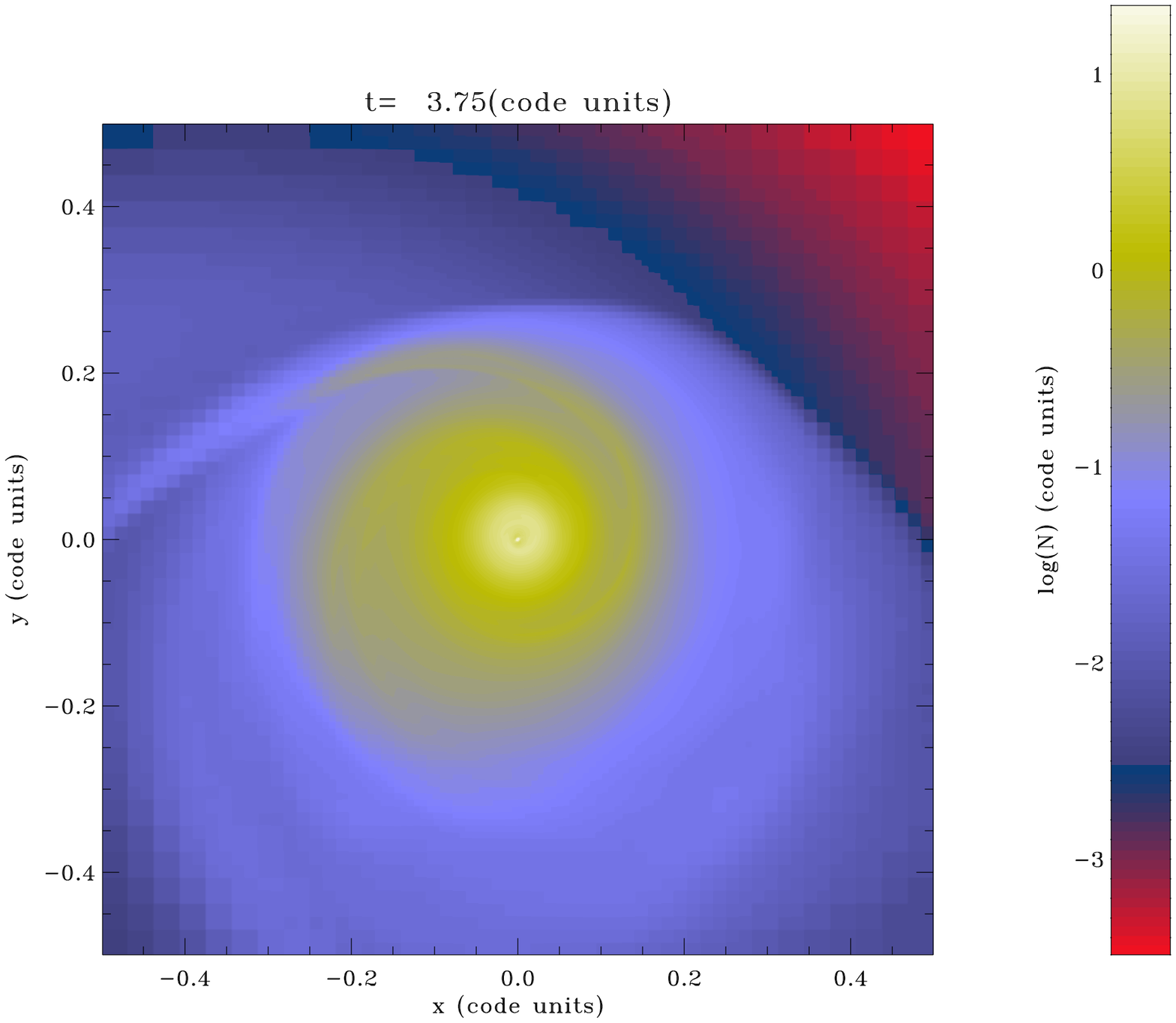}}  
\put(0,7){\includegraphics[width=8cm]{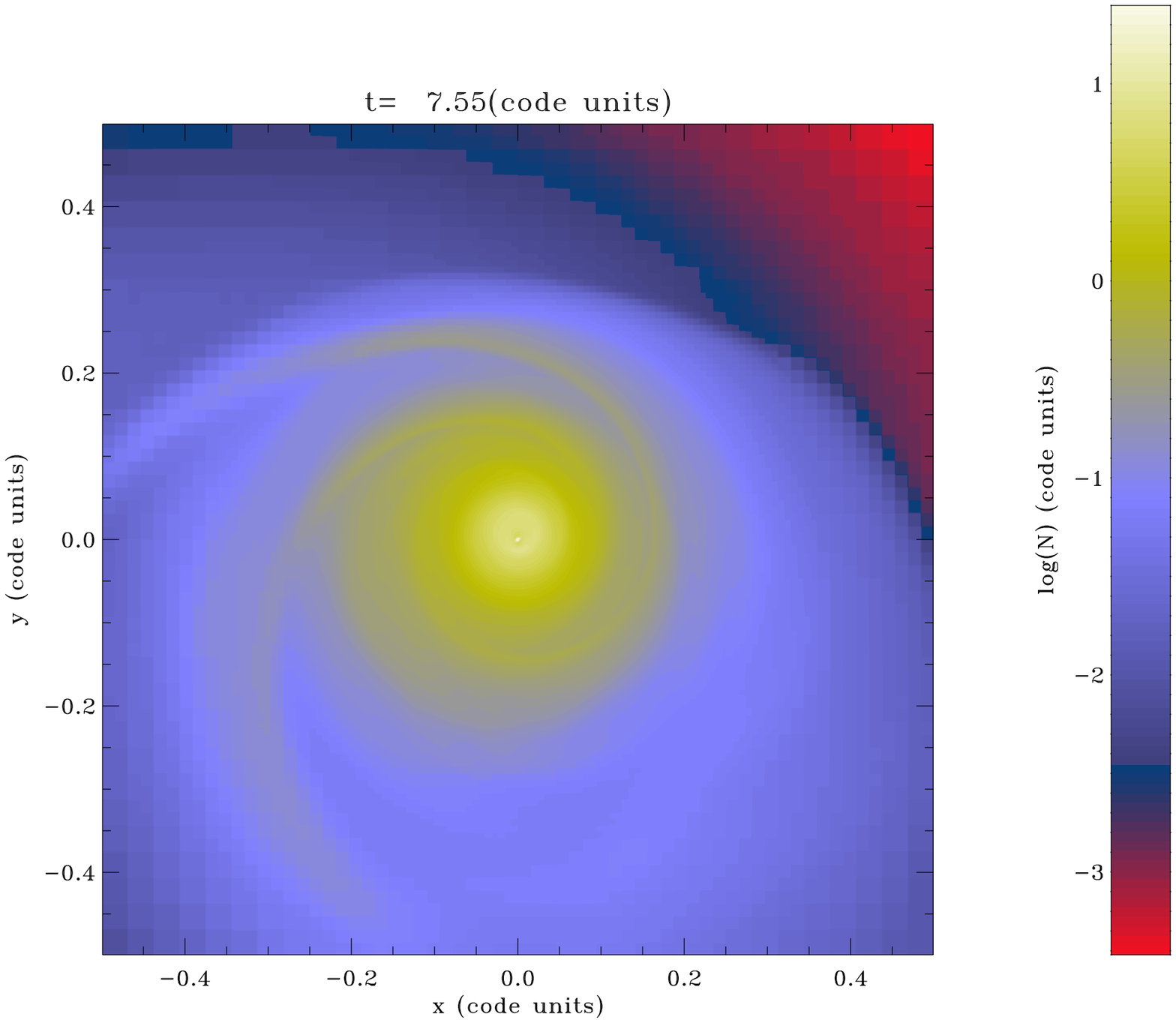}}  
\put(0,0){\includegraphics[width=8cm]{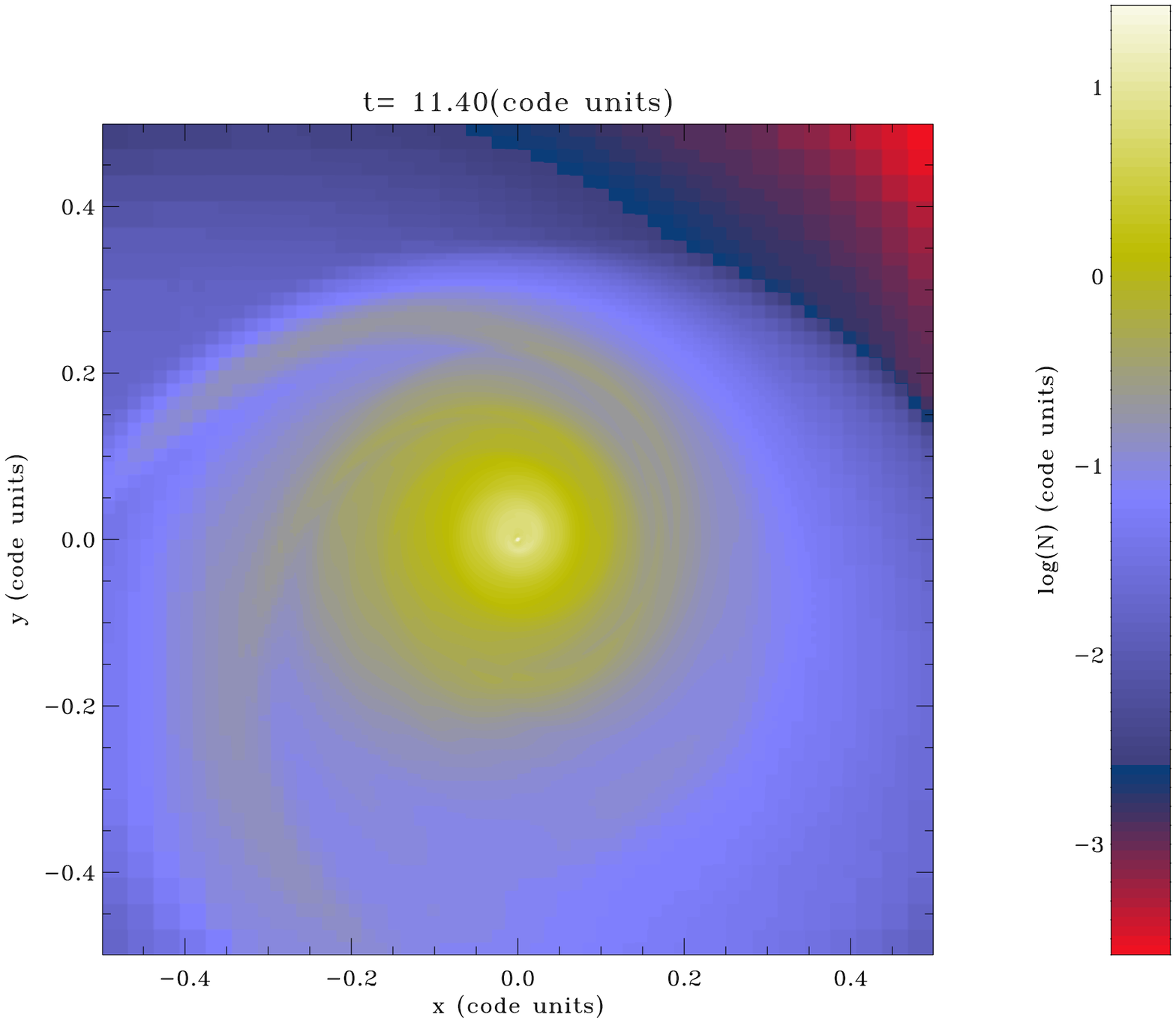}}  
\end{picture}
\caption{Simulation ASR2 (asymmetrical accretion with $U_0=2$ and standard resolution, left panels) and 
ASR2h (high resolution, right panels). Column density map
at 3 different timesteps. The disc initially has a radius of about $r=  0.25$. Note that the 
spiral patern is remarkably stationary.}
\label{col_map_2}
\end{figure*}

\subsection{The non-accreting case}

As it is important to understand the limits of the numerical experiment we  perform, we have first 
carried out a run without accretion. 
Figure~\ref{no_accret_map} shows a column density map at time 9.71 (recalling that the orbital time at the 
disc outer edge is about 0.75 this means that the disc has performed about 14 orbits). 
The disc remains smooth and symmetrical. There is a weak $m=2$ pattern located near the disc radius 
$r=0.25$. This is likely a consequence of the change of resolution there and the Cartesian mesh.
The resulting column density, which is displayed in the top panel of Fig.~\ref{no_accret} at 4 timesteps, 
does not evolve very significantly. 
It slightly diminishes in the inner part below $r<0.03$ because of the numerical resolution that becomes insufficient at small radius
(see discussion in Sect.~\ref{num_reso}). 
It also increases in the outer part at radii larger than $r>0.3$, that is to say close to the disc radius. This is expected since the 
gas is initially out of equilibrium there. 
Overall these effects remains fairly limited. We also computed the azimuthally averaged $\alpha$ parameter
defined as
\begin{eqnarray}
\alpha =    { < \rho \delta u_r \delta u _\phi>  \over < P > }.
\end{eqnarray}
where $\delta u_r$ and $\delta u_\phi$ are the fluctuation of radial and azimuthal velocities.
It is displayed in the bottom panel of Fig.~\ref{no_accret}. After  3-4 orbits, it is below $10^{-3}$ for radius 
$0.03<r<0.3$, that is to say inside the disc but sufficiently far from the 
center. After 8-10 orbits, it has dropped further reaching values close 
to $10^{-4}$ almost everywhere in the disc. At small radii, $r<0.03$, we see that $\alpha$ 
increases to large values. This is clearly a consequence of the insufficient numerical resolution and 
is consistent with the column density variations. 

We conclude that the numerical setup we use, permits us to investigate physical effects, which leads to 
transport characterized by $\alpha$ larger than 10$^{-4}$ or so in a region located between the 
disc radius, $r_d$, and about $r_d/5$. As reported below, this will indeed be the case 
for the choice of parameters, namely the external accretion rate, that will be considered.

\section{Results for non-axisymmetric  accretion of rotating gas}

\setlength{\unitlength}{1cm}
\begin{figure*} 
\begin{picture} (0,9)
\put(8,4.5){\includegraphics[width=9cm]{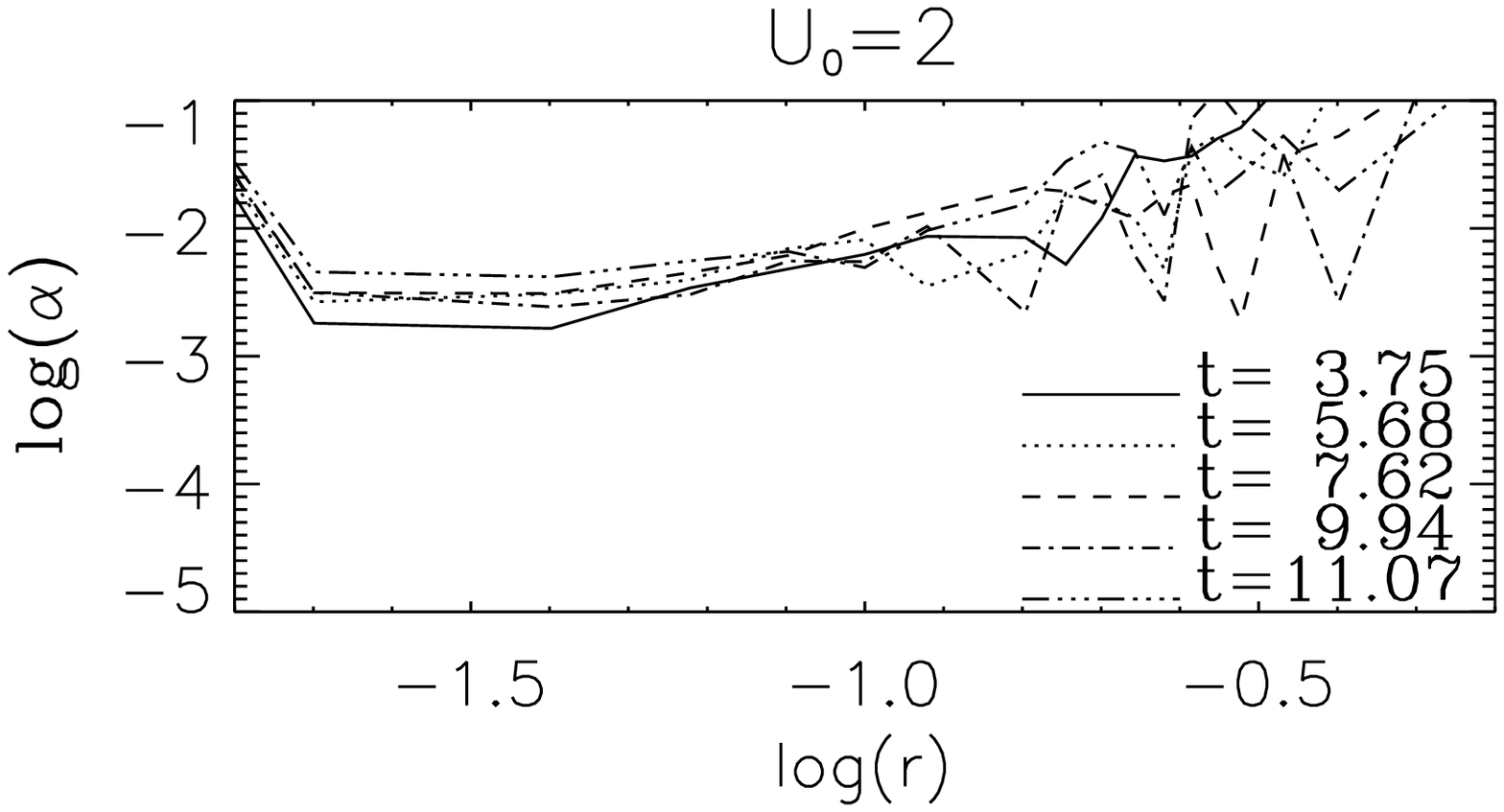}}  
\put(8,0){\includegraphics[width=9cm]{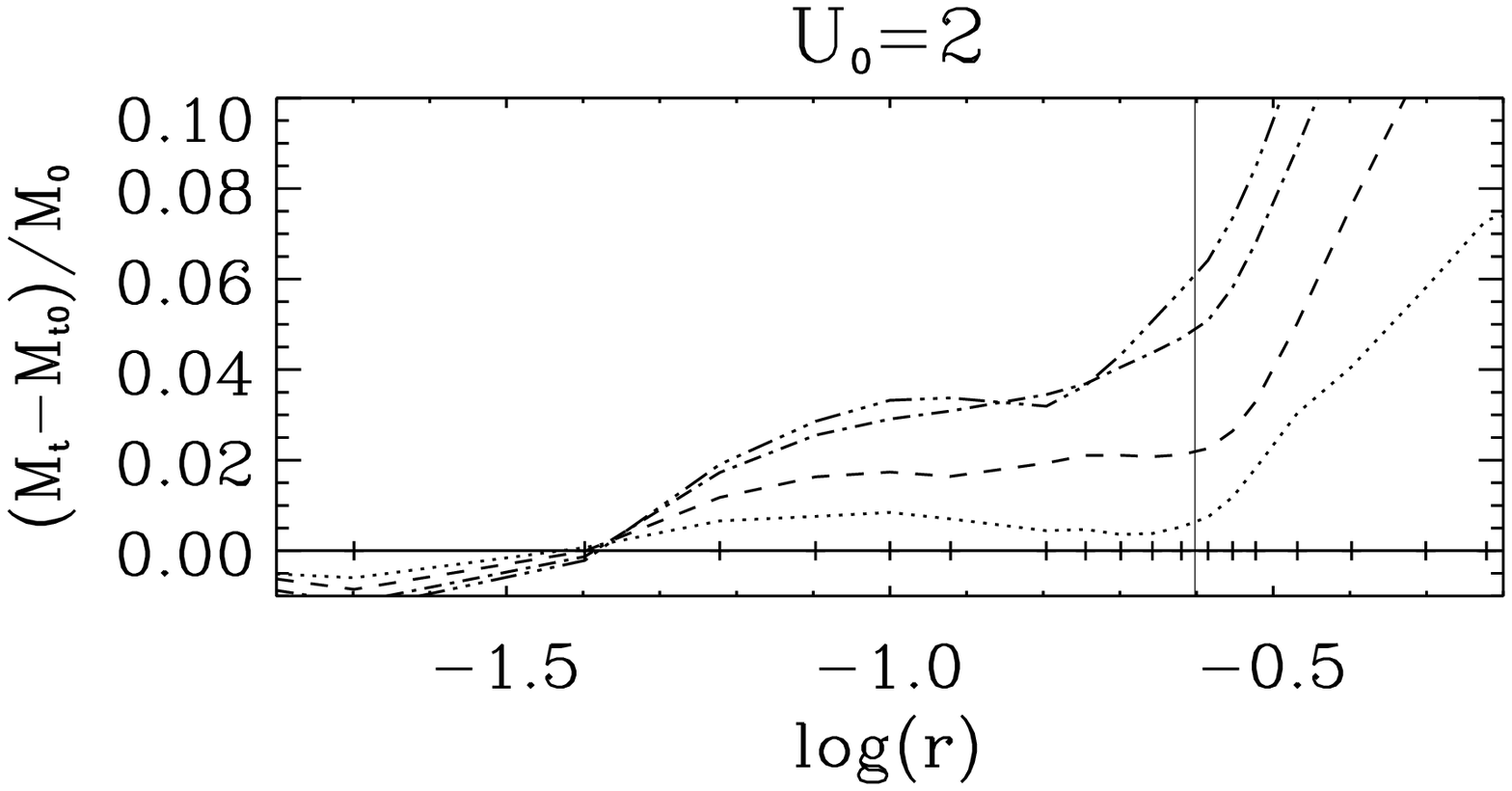}}  
\put(0,4.5){\includegraphics[width=9cm]{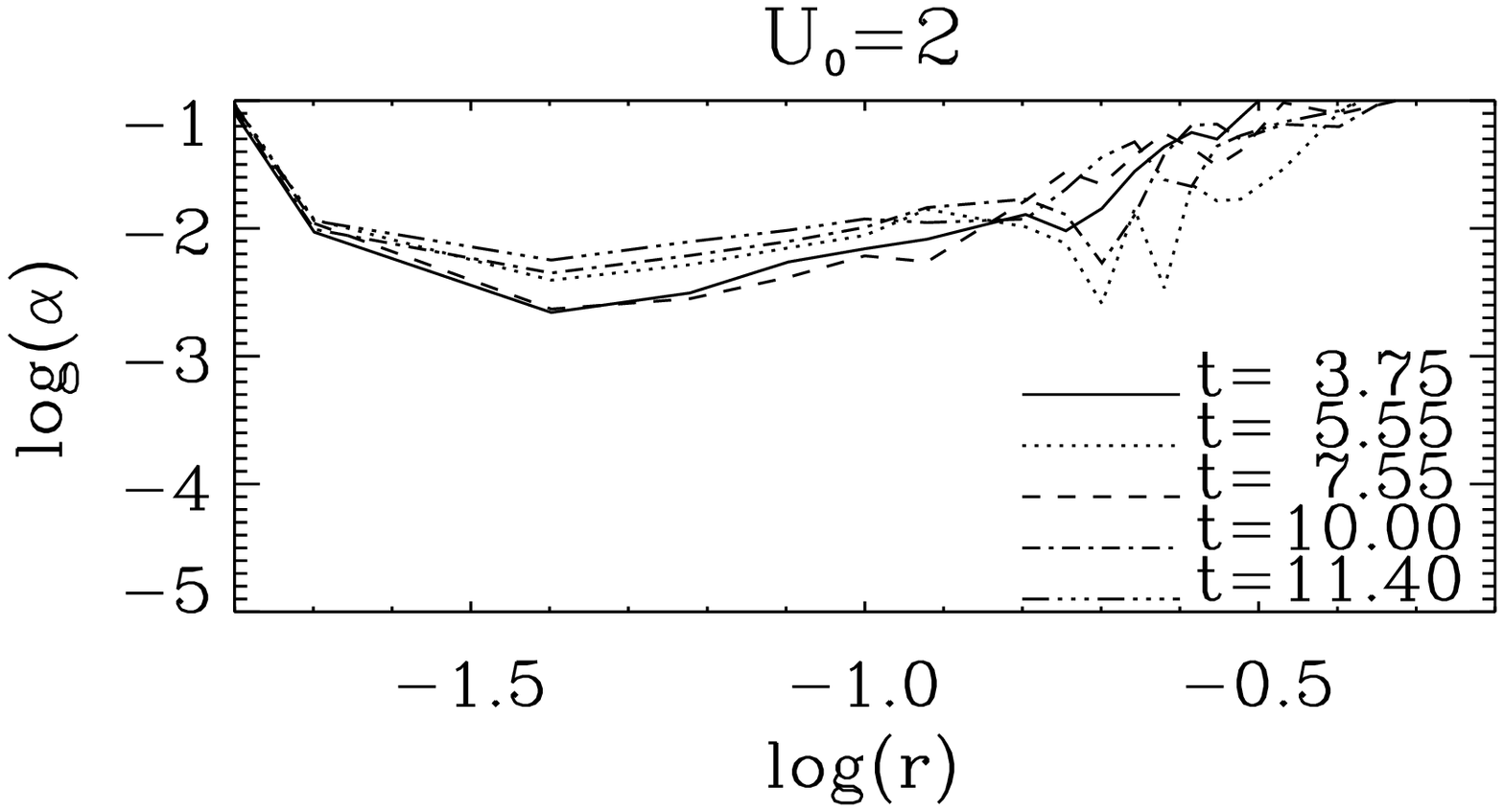}}  
\put(0,0){\includegraphics[width=9cm]{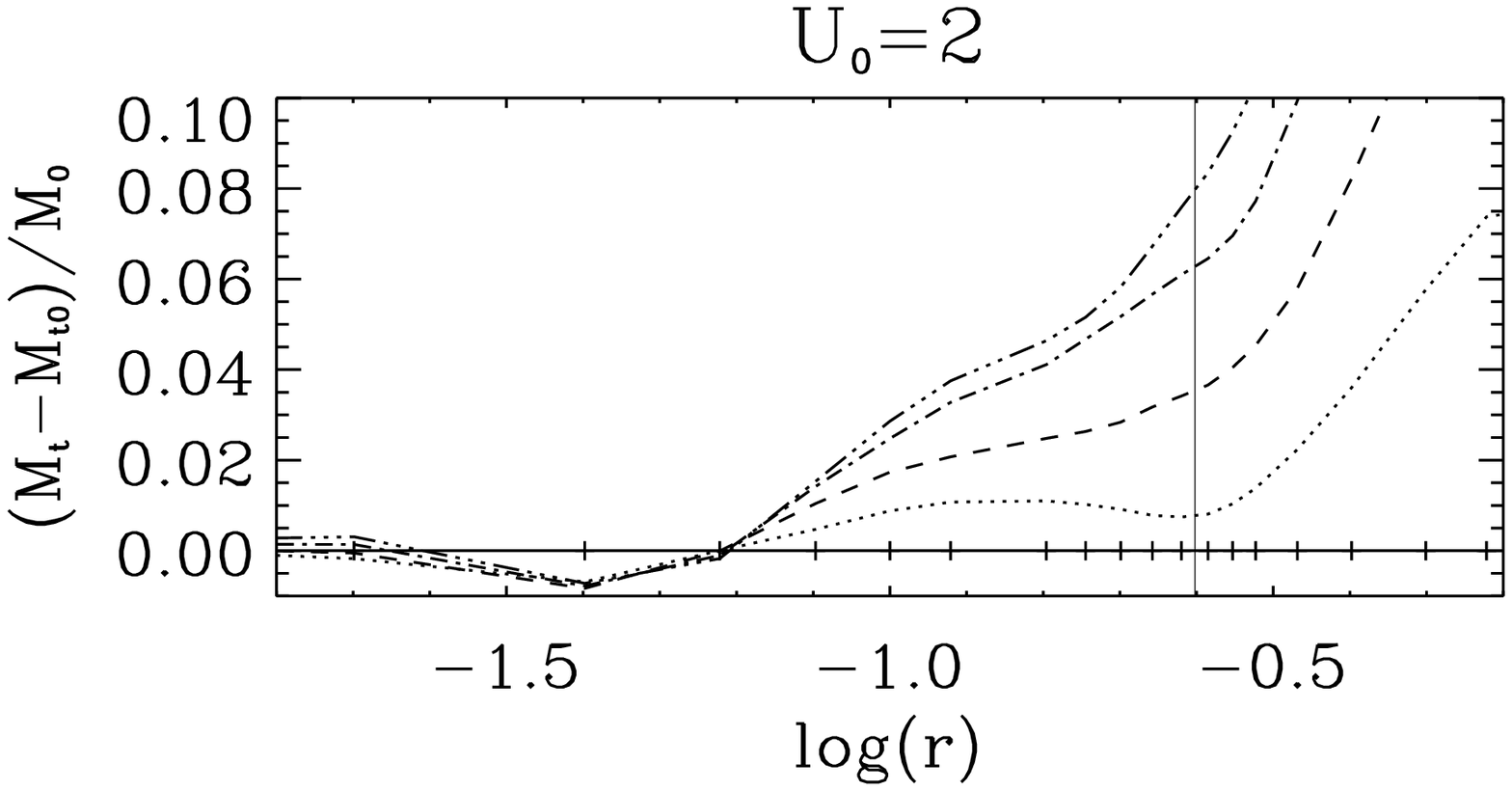}}  
\end{picture}
\caption{Simulations ASR2 (left) and ASR2h (right).
 Top panels: $\alpha$ 
as a function of the radius $r$. Bottom panels: cumulated mass difference over total mass (see text).
Solid lines at $r=0.25$ represent the initial disc edge. }
\label{alphcumul_fidu}
\end{figure*}


\subsection{Qualitative description}
In this section, we present in detail the simulation labelled as ASR2. It presents a non-asymmetric 
accretion flow, which contains the same amount of angular momentum as the gas at the disc outer edge. 
It has an accretion parameter, $P$, that is about $4 \times 10^{-2}$.
Figure~\ref{col_map_2} shows column density maps at 3 timesteps (from about 4 to 15 orbits) and at two 
different resolutions (left panels: standard resolution, right panels: high resolution).
Top panel shows various noticeable features. First of all, there is a prominent nearly stationary
spiral arm going from $x=-0.5$, $y=0$ to the disc. This is due to the incoming flow, which shocks 
onto the gas present within  the computational box (see further discussion in Sect.~\ref{discus}). This triggers the development of 
a prominent dense  spiral arm within the disc, starting at about $x=-0.2$ and $y=0.15$.
Second, there is a second spiral arm that is slightly less prominent and more or less symmetrical to the first 
one. Because of the gas distribution around the disc being not symmetrical, the whole disc is 
clearly not symmetrical indicating that $m=1$ perturbations have been developing. This is at stark 
contrast with the non-accreting disc presented in Fig.~\ref{no_accret_map}. 

The middle and bottom panels show later times. Clearly, the spiral patterns developed further 
and have propagated away reaching radius comparable to a few times the disc radius. 
This is reminiscent of the very prominent spiral pattern observed in numerical simulations 
of disc forming in class-0 protostars \citep[e.g.][]{joos+2012} and suggest that 
large scales spiral patterns are a natural outcome of externally accreting discs. 

\subsection{Accreted mass and  $\alpha$ parameter}
To quantify the transport of mass, that results from the large asymmetry induced by the 
incoming flow, we have computed the cumulated mass through the disc at various timesteps
as well as the mean $\alpha$ parameter. 

Bottom panel of Fig.~\ref{alphcumul_fidu} shows the cumulated mass (that is to say the mass contained within a cylinder
of radius $r$)  difference between a reference time (first timestep quoted in the figure) and 
4 later timesteps, divided by the total mass within the box at the reference time (which is close to the mass within the disc). 
The four curves show similar features. First, the cumulated mass 
rapidly increases with radius in the region located outside the disc radius, 
$r > r_d$. There is a clear break located approximately at the disc radius, 
$r_d \simeq 0.25$ below which the cumulated mass varies less stiffly with 
radius. At $\log(r) \simeq -1.2$, the cumulated mass remains constant with
 time, implying that no mass is crossing this region. Below this radius, 
the cumulated mass difference slightly decreases with time implying that the 
gas there is getting transported outwards (but not further than $\log(r) \simeq -1.2$ where the difference in cumulated mass
vanishes). This limit likely represents the 
point above which the simulation results are dominated by grid effects as anticipated 
in Sect.~\ref{num_reso}.

The $\alpha$ parameter is plotted in the top panel of Fig.~\ref{alphcumul_fidu}. The general behaviour is 
similar to the no-accretion case (Fig.~\ref{no_accret}), $\alpha$  is large in the very inner part, because 
of insufficient resolution, and in the disc outer part,  where the external material is falling in. 
At intermediate radius, there is  however a plateau with values that increase with time 
to reach about $10^{-2}$. Since this value is about 30 times larger than the values obtained 
in the non-accretion case, this clearly shows that infall is responsible for the 
measured $\alpha$.

\subsection{The issue of numerical convergence}


To verify the reliability of these simulations, we have performed higher resolution simulations.  
The right panels of Fig.~\ref{col_map_2} display the column density map at 3 timesteps.
 Very similar pattern can be observed for the 2 resolutions, seemingly indicating 
that the simulations are not dominated by insufficient resolution effects inside the disc. 

More quantitative comparisons are given in Fig.~\ref{alphcumul_fidu}  where the right panels
present the high  resolution run (ASR2h).  
The values of $\alpha$ displayed are   similar although run ASR2 indicates slightly larger $\alpha$ 
at radius, $r< 0.1$.
The cumulated mass shown at  the bottom panels of Fig.~\ref{alphcumul_fidu} is very similar in the 
outer part, that is outside the disc ($r > 0.25$). The profile are however more different inside the 
disc. In particular, the cumulated mass vanishes at smaller radii for ASR2h than for ASR2. The profile 
of the former is also flatter than the one of the latter. This is clearly an impact of the resolution
and this suggests that the limited resolution prevents accretion up to the center.

Altogether these results suggest that accretion triggered by the external flow is able to penetrate 
significantly inside the disc probably at least down to a radius equal to $\simeq r_d / 5$, this number reflecting however the 
limit set by the highest resolution we could achieve. 
 Typically,  we find that with the present configuration, roughly 11$\%$ of the mass that has been injected in the box is accreted inside the disc
at a radius below $r_d/2$. At a radius below  $r_d$, this  number can be larger than 20$\%$, though defining the disc boundary is arbitrary because of the 
complexity of the flow in this region. This indicates that roughly 50$\%$ of the mass that is captured by the disc is able to penetrate at 
radius smaller than $r_d/2$.

\subsection{Comparison between analytical and numerical solutions}

\setlength{\unitlength}{1cm}
\begin{figure} 
\begin{picture} (0,12)
\put(0,8){\includegraphics[width=9cm]{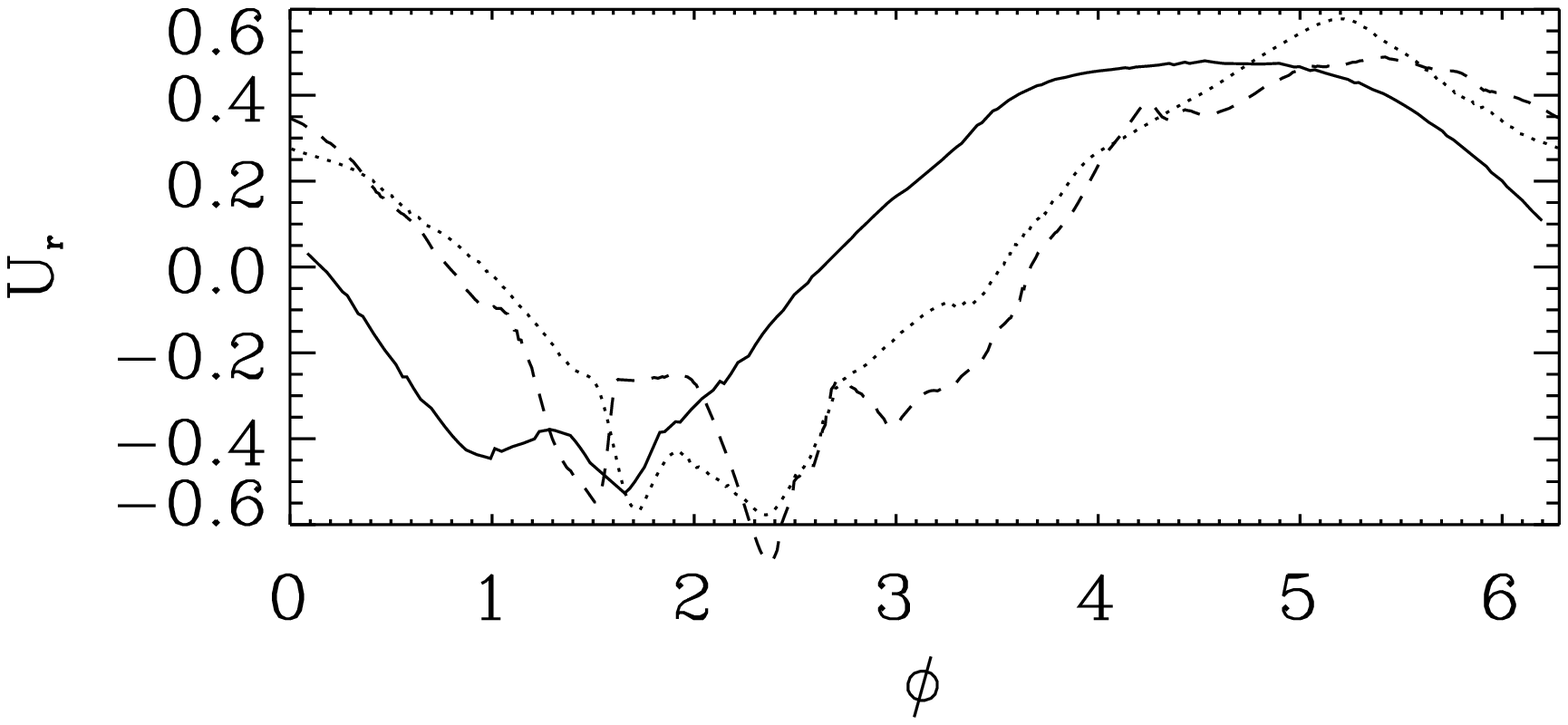}}  
\put(0,4){\includegraphics[width=9cm]{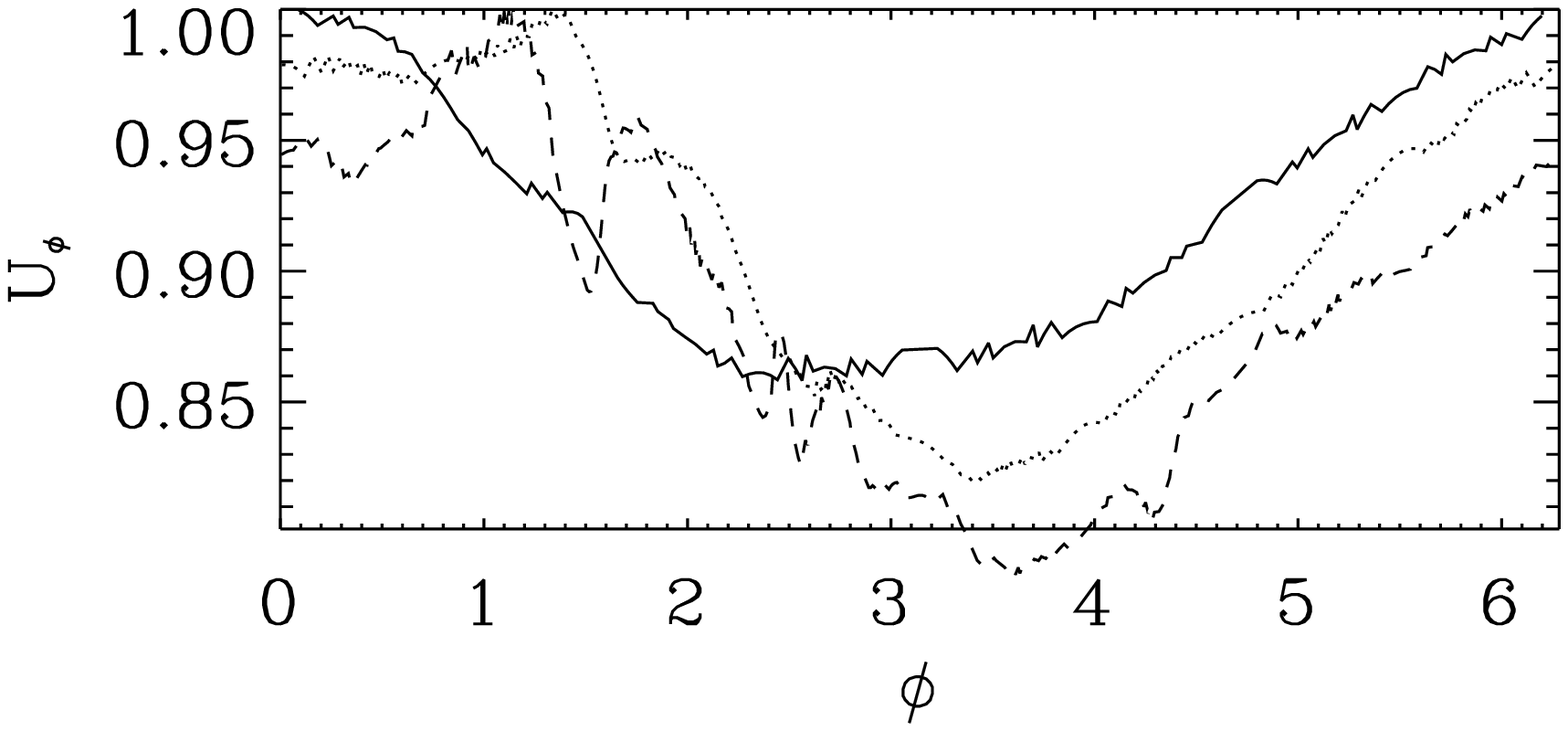}}  
\put(0,0){\includegraphics[width=9cm]{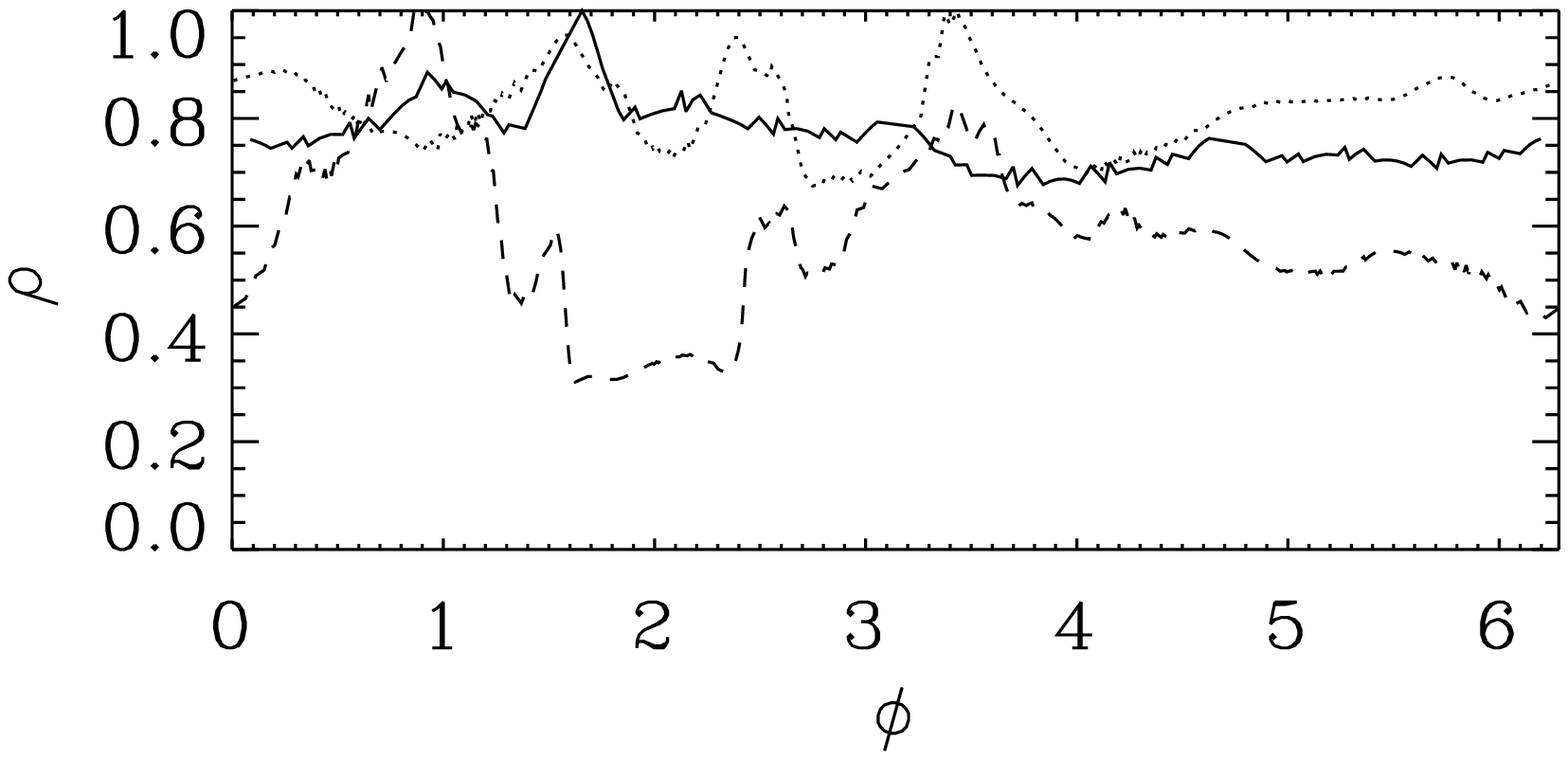}}  
\end{picture}
\caption{Azimuthal profiles from simulation ASR2h at time $t=0.75$ and at radius 
$r=$ 0.08 (solid line), 0.14 (dotted), 0.18 (dashed). Note that the density and 
the azimuthal velocity have been normalised  such  that their peak value 
is equal to 1.}
\label{azimuth_prof}
\end{figure}

In an attempt to better understand the physical mechanism responsible for the transport 
of mass and angular momentum in the disc through the spiral driven accretion, self-similar 
solutions in the $r$ variables, have been studied in \citet{hennebelle+2016} following 
the approach of \citet{spruit87}. 
These exact solutions
of the inviscid fluid equations depend on the $\phi$ variable, the polar angle,
 and present two essentials features. First of all they undergo shocks which 
play an essential role since this is where energy is  dissipated, an essential feature for a system 
that transports angular momentum. Second, the radial velocity is changing sign with $\phi$. This 
implies that there are both inward and outward mass fluxes. In particular the latter is responsible 
for the outward flux of angular momentum. 

While self-similar solutions remain restricted by the heavy assumptions regarding their dependence, 
these two aspects constitute essential predictions that should be generic for any accretion discs
driven by spiral waves, particularly in the absence of other forces such as self-gravity and magnetic field, 
which could exert a torque on the gas. To verify whether these features are indeed playing a role 
in the present simulations, Fig.~\ref{azimuth_prof} shows the azimuthal profiles of $U_r$, $U_\phi$ and $\rho$ at 
three different radii. As can be seen the profiles present some similarities with the self-similar solutions. 
In particular, the radial velocity, $U_r$ is clearly changing sign. It remains negative in about one third 
of the angular distance and positive for the rest. Moreover there are clear discontinuities, seen 
both in $U_r$ and $U_\phi$ which are similar to the shocks of the self-similar solutions (see Fig.~1 of \citet{hennebelle+2016}).
Unsurprisingly, there are also significant differences. First, the dominant mode is 
the $m=1$ one, a clear consequence of the accretion pattern, which is also $m=1$. No analytical solution was found 
for the $m=1$ case. Second, there are multiple shock features instead of one. 
Finally, the density fields present more structure in the numerical simulations than in the analytical solutions. 
These differences are likely consequences of the unstationnary nature of the numerical solutions.

\section{Dependence on infall rate and comparison with bidimensional simulations}

\subsection{Dependence on infall rate}

\setlength{\unitlength}{1cm}
\begin{figure} 
\begin{picture} (0,9)
\put(0,4.5){\includegraphics[width=9cm]{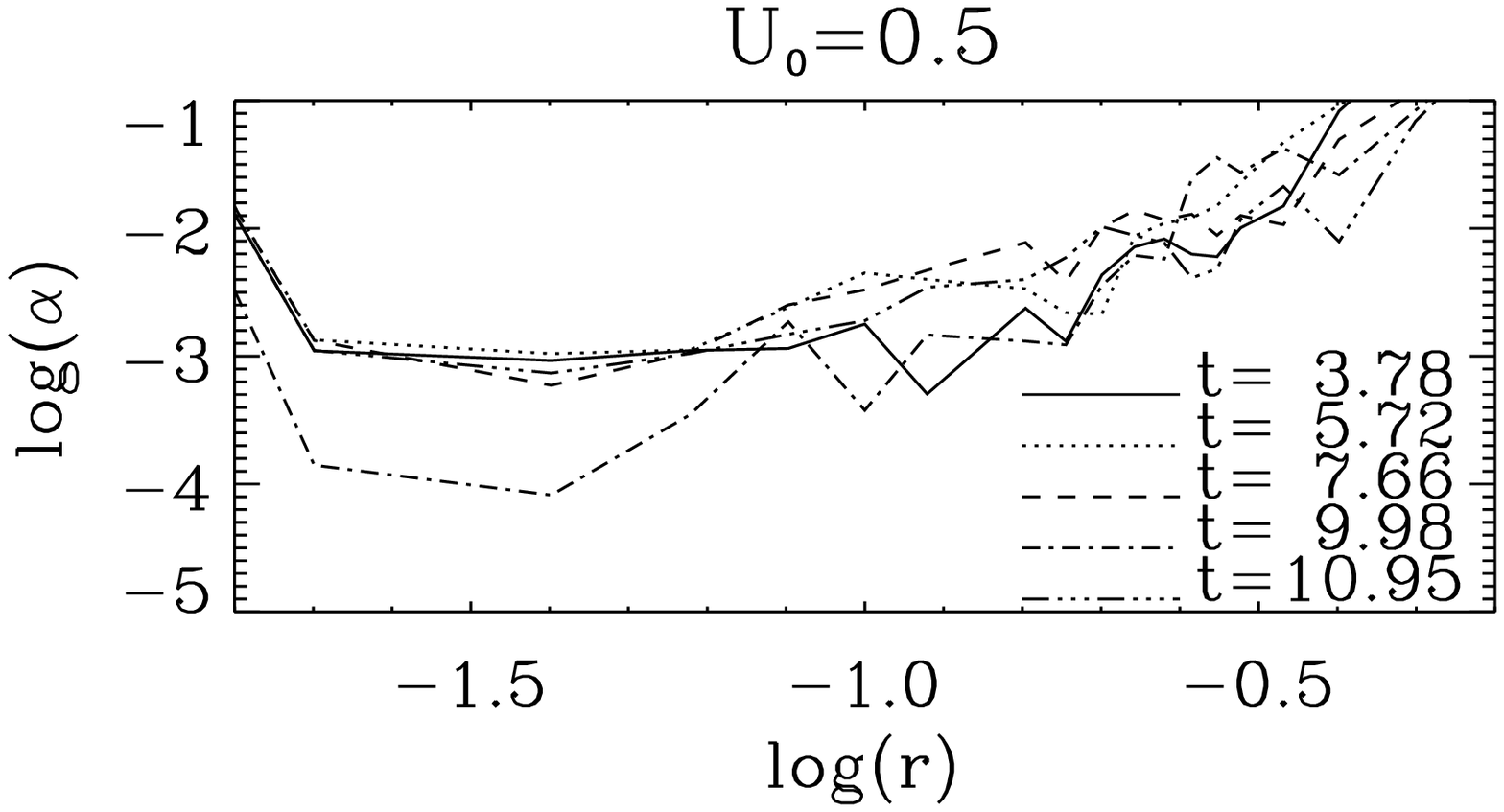}}  
\put(0,0){\includegraphics[width=9cm]{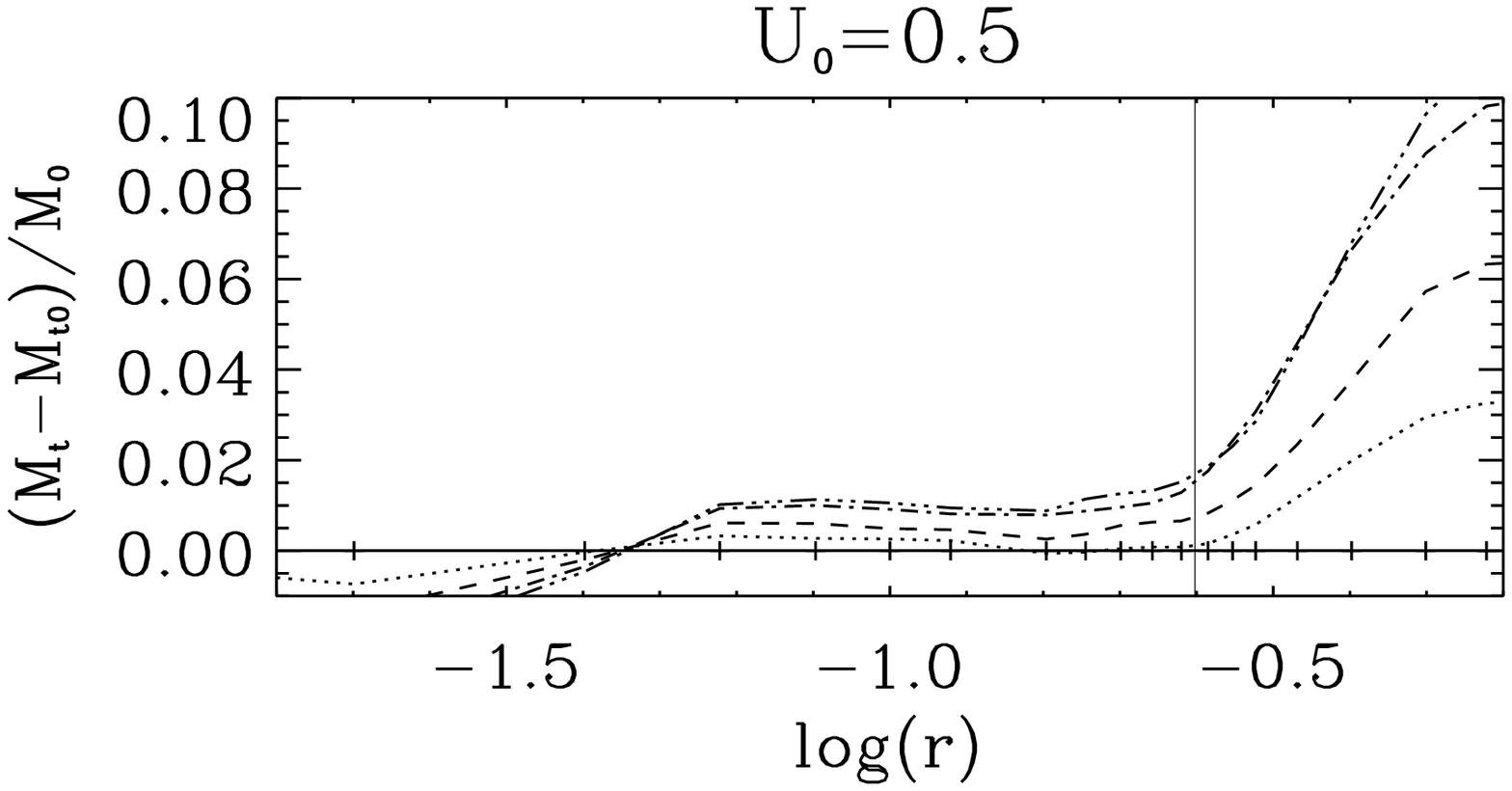}}  
\end{picture}
\caption{Simulations ASR0.5h. $\alpha$ values and cumulated mass difference as a function of radius, $r$.}
\label{alpha_0.5}
\end{figure}

To investigate the influence that infall has on the disc dynamics, we show a smaller accretion rate 
characterized by an infall velocity equal to $U_0=0.5$. As shown by Eqs.~(\ref{P_ttauri})-(\ref{P_code})
this corresponds to an accretion rate inside the computation box of about 10$^{-7}$ M$_\odot$ yr$^{-1}$.
The effective accretion rate for the disc  is $\simeq$5 times smaller since as already explained most of the mass 
does not fall into the disc and remains in the surrounding region.
The corresponding $\alpha$ 
profile and cumulative mass distribution are shown in Fig.~\ref{alpha_0.5}. The values of $\alpha$ 
vary from a few 10$^{-3}$ at $\log r = -0.9$ to about 10$^{-3}$ at $\log r = -1.3$ at time 10.95. Note that these 
values fluctuate in particular at time 9.98 where they are typically 2 times smaller (for $\log r > -1.3$). For comparison 
the $\alpha$ values  for $U_0=2$ are about 4 times larger with values on the order of $5 \times 10^{-3}$ at 
$\log r = -1.3$ and almost $10^{-2}$ at $\log r = -0.9$.

As expected, the cumulated mass is also lower with $U_0=0.5$. Its value at time 10.95 and at radii $\log r=-0.9$
is about $1.5 \times 10^{-3}$ while at similar time and radius it is about $5.5 \times 10^{-3}$ for 
$U_0=4$, that is to say roughly 4 times larger. 
The mass accreted inside the disc (at $\log r<-1$) is about 20\% of the total mass
injected in the computational box. Interestingly, we see that the profile of the cumulated mass is 
fairly flat indicating that most of the mass accreted within the disc is transferred inwards.


\setlength{\unitlength}{1cm}
\begin{figure} 
\begin{picture} (0,9)
\put(0,4.5){\includegraphics[width=9cm]{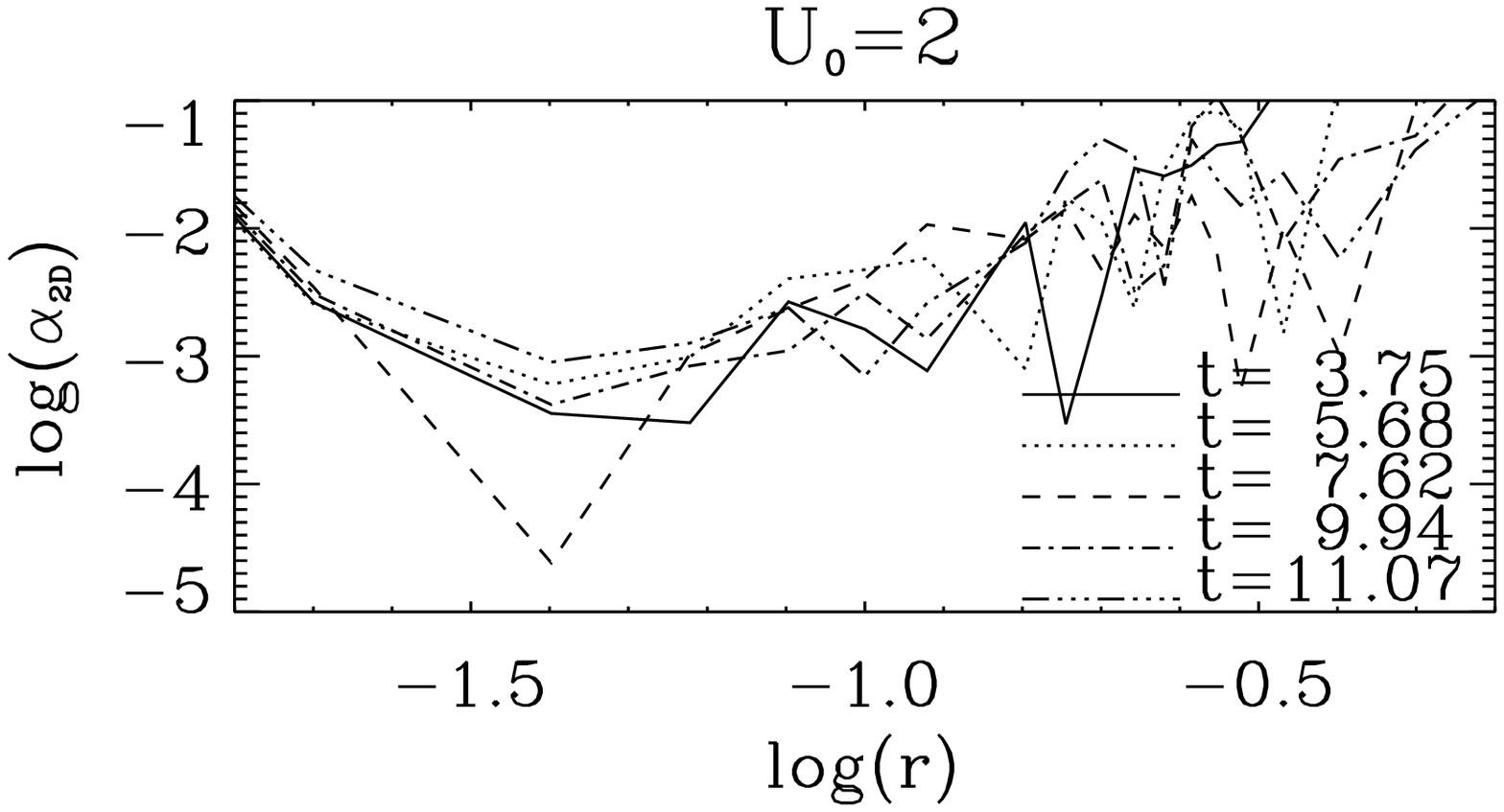}}  
\put(0,0){\includegraphics[width=9cm]{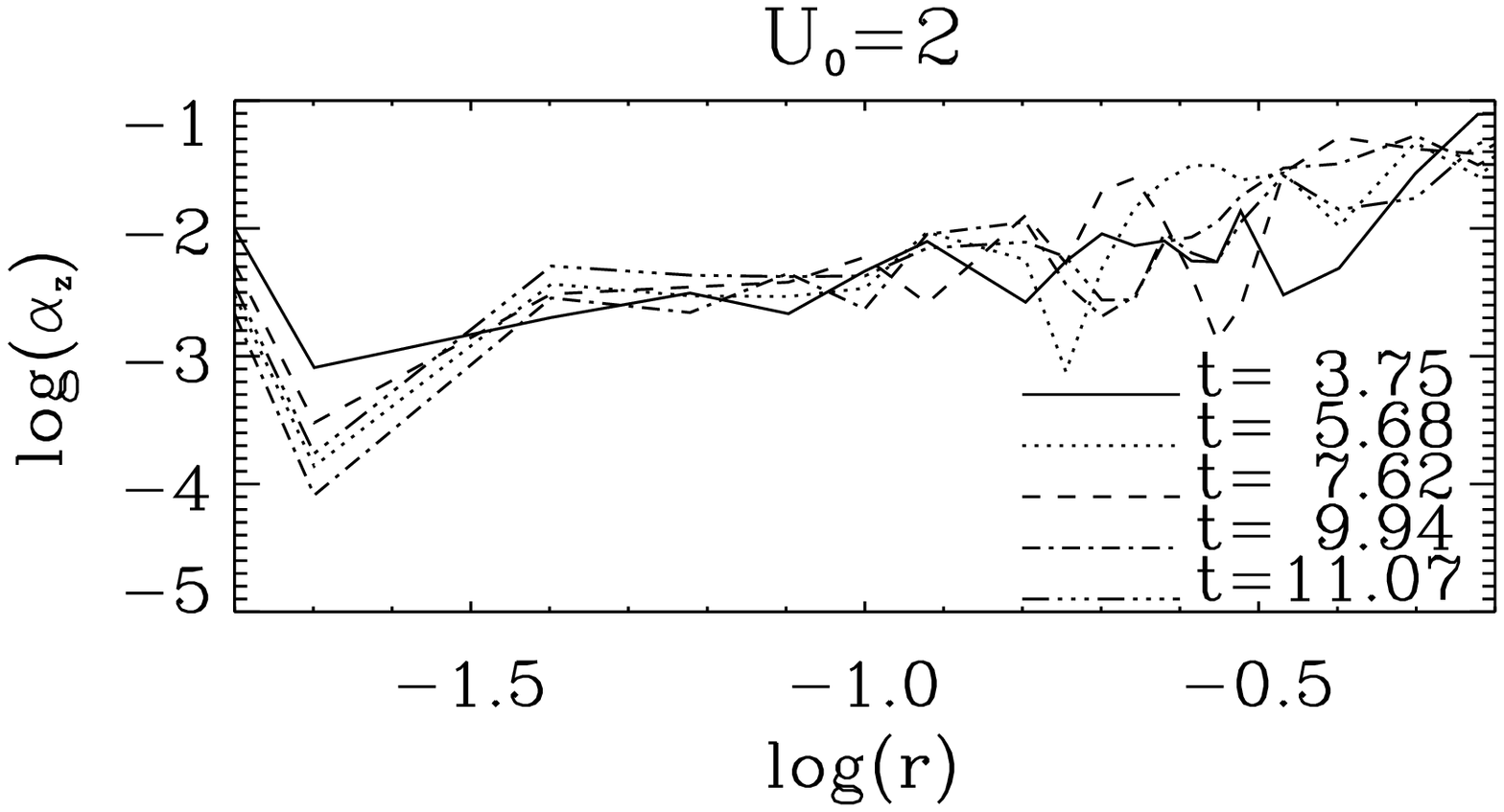}}  
\end{picture}
\caption{Values of $\alpha _{2D}$ (obtained by averaging all quantities along the z-axis) and $\alpha _z$ (the difference of $\alpha$ and 
$\alpha_{2D}$ representing the contribution of the fluctuations along z) for run ASR2h. The two quantities have comparable values in the 
inner part of the disc.}
\label{alpha_z}
\end{figure}

\subsection{Comparison with the 2D simulations}
In their 2D simulations, \citet{lesur+2015} estimate that for an accretion rate of 
$10^{-7}$ M$_\odot$ yr$^{-1}$, the asymmetrical case with rotation, leads to 
values of $\alpha$ of about $4 \times 10^{-4}$ which is therefore 2-3 times lower than the values 
obtained in our 3D simulations. The same is true for the simulations ASR2h, which corresponds
to accretion rate of about $4 \times 10^{-7}$ M$_\odot$ yr$^{-1}$, as in the 2D case 
$\alpha$ values slightly below 10$^{-3}$ would be expected (from Fig.~4 of \citet{lesur+2015}).
Therefore  the 3D simulations performed here present $\alpha$ values about 3 times larger than 
in the 2D simulations. Further corrections must be taken into account, however. Indeed,  
\citet{lesur+2015} used an aspect ratio, $H/R$ of 0.1 while the one employed here is 0.15, since 
from analytical analysis
$\alpha \propto (H/R)^{1.5}$ \citep{hennebelle+2016}, the $\alpha$ values in the 3D case may be larger than in the 2D one 
by a factor of about 1.5-1.6.

\subsubsection{Quantifying the importance of 3D effects}
To investigate why the  typical $\alpha$ values are larger in 3D than in 2D, we have 
decomposed $\alpha$ in two contributions which represents its bidimensional contribution 
($\alpha_{2D}$) and its remaining 3D  contribution due to variation along the z-axis ($\alpha_z$). To achieve this, 
we proceed as follows. First, we take the z-averaged quantities defined for example as 
\begin{eqnarray}
\bar{\ u_r} =    {\int \rho u_r dz  \over \int \rho dz },
\end{eqnarray}
Then we can simply calculate an $\alpha$ in 2D as:
\begin{eqnarray}
\alpha _{2D} =    { < \bar{\rho} \bar{\delta u_r} \bar{\delta u _\phi} >  \over < \bar{P} > }.
\end{eqnarray}
where the bracket means the averaged values over the  radius and azimuth.

Finally, it is easy to show that $\alpha _z= \alpha - \alpha_{2D}$ is equal to
\begin{eqnarray}
\alpha _{z} =    { < \rho (u_r - \bar{u_r})  (u _\phi - \bar{u_\phi}) >  \over < P > }.
\end{eqnarray}
where the bracket represents the averaged values over radius, azimuth and z-coordinates.

In practice,  this is achieved by selecting all cells within bins of radius and 
azimuthal coordinates. 
The resulting values of $\alpha_{2D}$ and $\alpha_z$ are displayed in Fig.~\ref{alpha_z}. As can be seen while $\alpha_{2D}$ dominates
in the outer part of the disc, they are comparable in the inner part ($log r < -1$). There is a possible trend of 
$\alpha_z$ becoming dominant over $\alpha_{2D}$ at small radii.

\subsubsection{Physical interpretation}

\setlength{\unitlength}{1cm}
\begin{figure} 
\begin{picture} (0,7)
\put(0,0){\includegraphics[width=8cm]{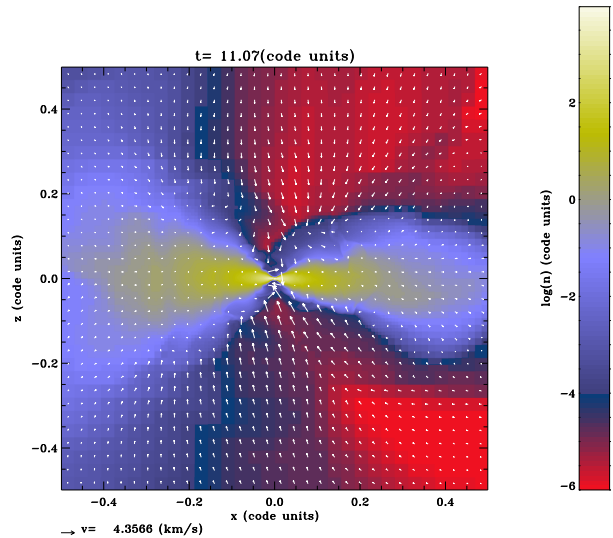}}  
\end{picture}
\caption{Density and velocity field in the xz plane of the  simulation ASR2.
As can be seen accretion is also proceeding from above and below the disc plane along the 
z-axis.}
\label{dens_xz}
\end{figure}

Since it appears that structures along the z-axis are playing a role in producing somewhat larger 
values of $\alpha$, Fig.~\ref{dens_xz} displays the density and velocity fields  in the xz plane 
for the ASR2h  simulations. The result is as expected, a significant velocity field is seen 
in the z-direction and the shape of the disc appears to be irregular and not symmetric.   
We believe that this latter point is at the origin of the observed difference. Indeed in 3D the gas 
can enter through the disc directly from above and below instead of inflowing inside the 
equatorial plane only as in 2D. This certainly produces a higher accretion rate but also is 
likely to generate more turbulence through accretion since gas that approaches the disc from 
above undergoes a violent gravitational acceleration along the z-axis and shocks. This infall process
likely drives  further turbulence.

This could indicate that 2D simulations may generically  underestimate 
the accretion driven turbulence process in accretion discs. Indeed unless the  gas temperature 
varies abruptly inside the disc,  the scale height 
of the accreting gas must be, at least equal, to that  of the disc at its outer edge. 
Thus as the accreting gas is expected to fall onto the disc supersonically, it seems clear 
that even if approaching the disc close to the equatorial plane, it is not going to pile at the 
outer disc radius but likely will reach smaller radii from above (bottom).

\section{Influence of symmetry and angular momentum}
In this last section we briefly explore the impact of other aspect of the 
accretion flow, namely its symmetry and its angular momentum. 
Indeed, it is important to remember that the exact way accretion proceeds is not 
well known as it depends on the whole history of the star formation process. 
Previous 
2D simulations carried out in \citet{lesur+2015} have shown that indeed 
the accretion rate is not the only parameter to have a significant 
influence on the disc.

\setlength{\unitlength}{1cm}
\begin{figure} 
\begin{picture} (0,14)
\put(0,7){\includegraphics[width=8cm]{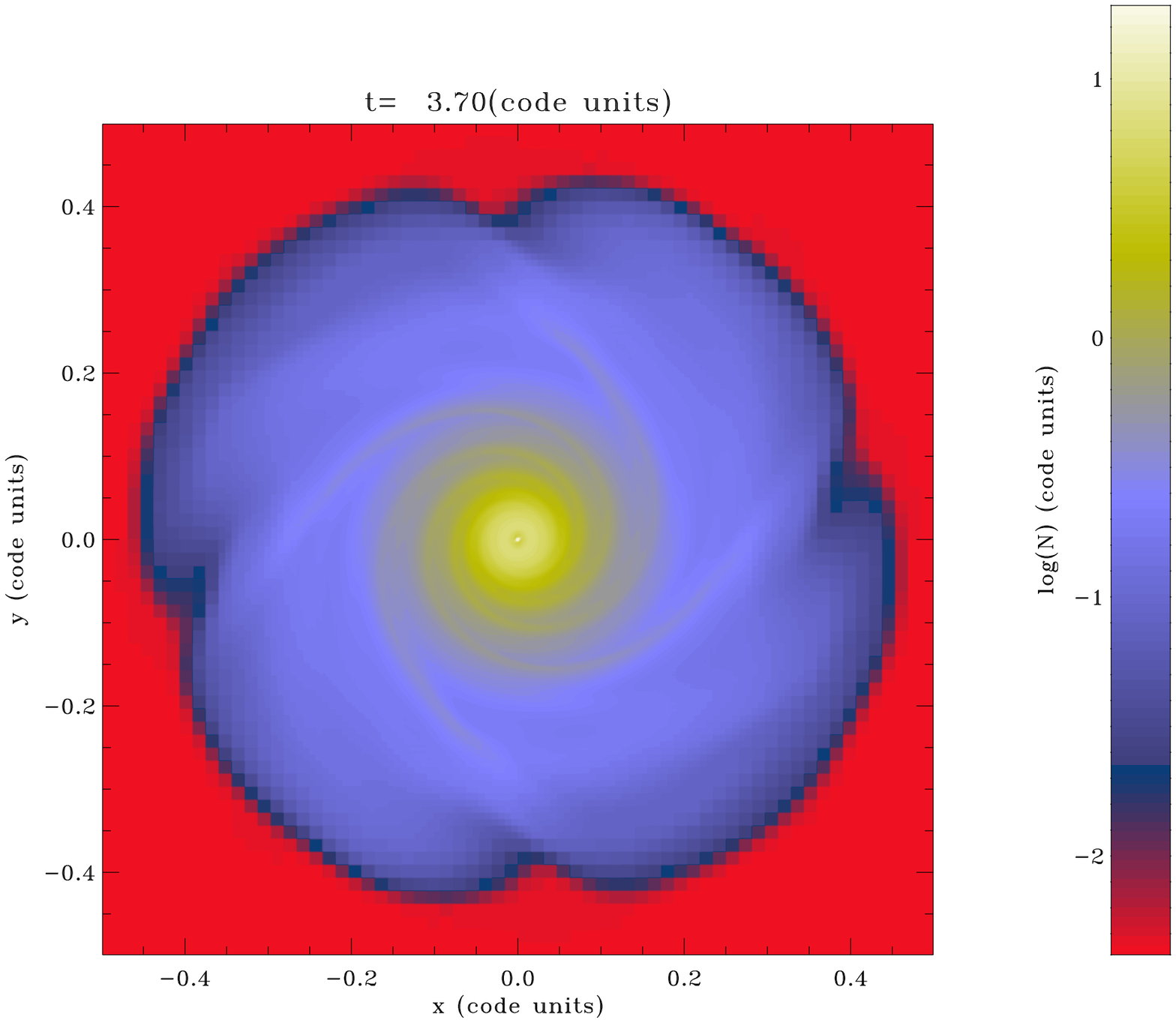}}  
\put(0,0){\includegraphics[width=8cm]{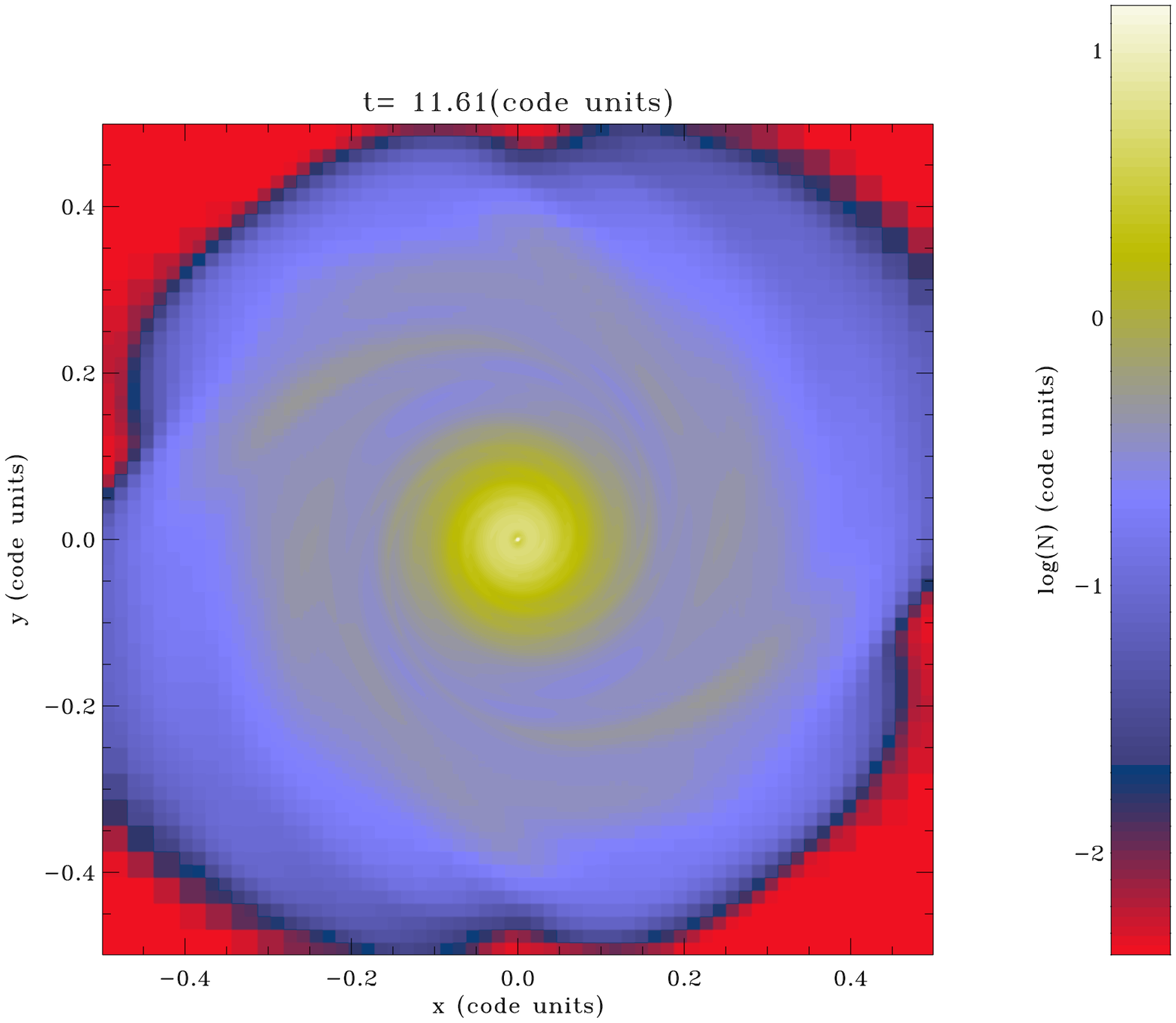}}  
\end{picture}
\caption{Simulation SR2 (symmetrical accretion with $U_0=2$). Column density map
at 2 different timesteps.}
\label{accret_sym_map}
\end{figure}

\setlength{\unitlength}{1cm}
\begin{figure} 
\begin{picture} (0,9)
\put(0,4.5){\includegraphics[width=9cm]{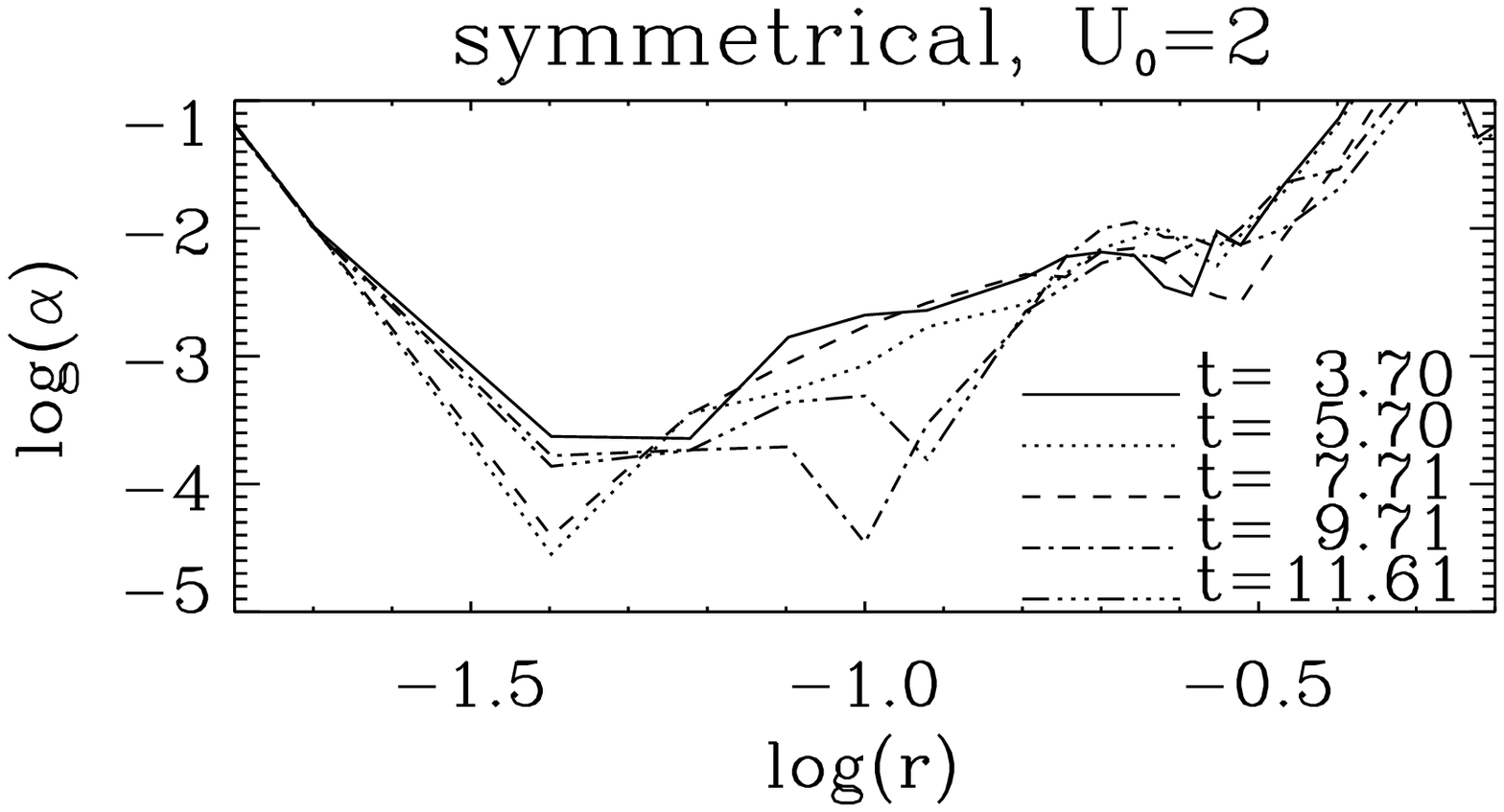}}  
\put(0,0){\includegraphics[width=9cm]{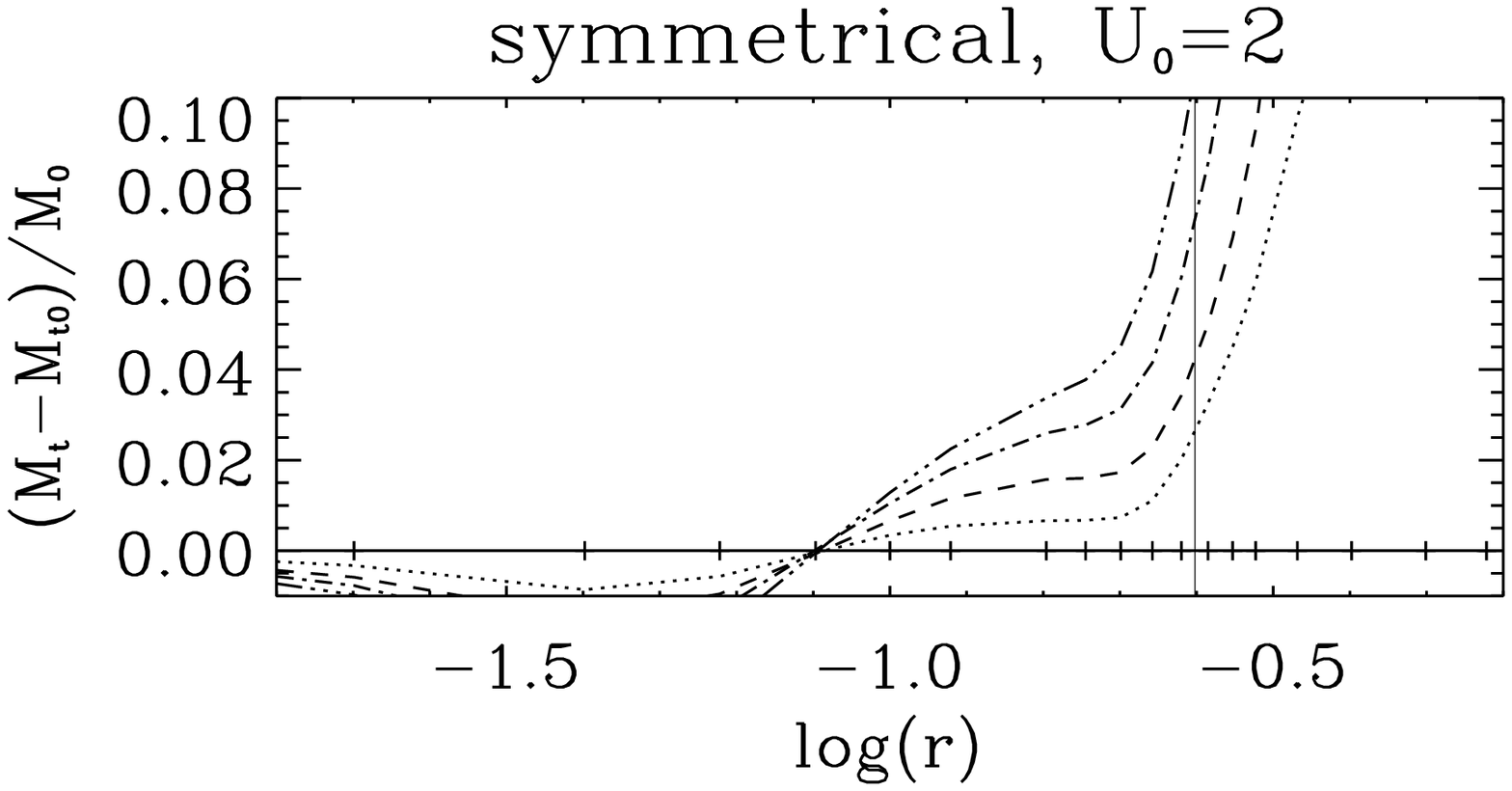}}  
\end{picture}
\caption{Symmetrical accretion rate with $U_0=2$. 
$\alpha$ values and cumulated mass difference as a function of radius, $r$.
}
\label{alpha_sym}
\end{figure}

\subsection{The axisymmetric case}

Figure~\ref{accret_sym_map} shows the column density at 2 different timesteps for simulation SR2. 
It clearly appears that the dominant mode is not the $m=1$ ones, as  is the case 
for ASR2. Instead the $m=4$ has developed and became dominant. 
Note that Cartesian coordinates are likely responsible for the excitation of the $m=4$ mode.
In the 2D axisymmetric simulations presented in \citet{lesur+2015}, the $m=4$ mode did not 
 dominate.
The spiral wave is tightly wounded in the inner part of the disc and more open 
in the outer part. Let us remind that in \citet{hennebelle+2016}, it was shown that 
the mass and angular momentum fluxes are higher for smaller $m$. This is a simple consequence
of the disc being more axisymmetric for high $m$.

The $\alpha$ value and the cumulative mass distribution are shown in Fig.~\ref{alpha_sym},
 they should be compared with simulation ASR2, which has the same resolution (left panel of 
Fig.~\ref{alphcumul_fidu}). It appears that while the $\alpha$ values are 
comparable in the outer part of the disc ($\log r \simeq -0.6$) to what is obtained 
for ASR2, they quickly drop in the inner part of the disc reaching 
almost the floor values inferred from the non-accreting simulation.

This behaviour is in good agreement with the cumulative mass distribution. 
Indeed while it is  as high as in ASR2, in the outer part of the disc
it quickly drops in the disc inner part indicating that the mass flux is significantly 
lower than for ASR2.

\subsection{Accretion of  non-rotating gas}

\setlength{\unitlength}{1cm}
\begin{figure} 
\begin{picture} (0,14)
\put(0,7){\includegraphics[width=8cm]{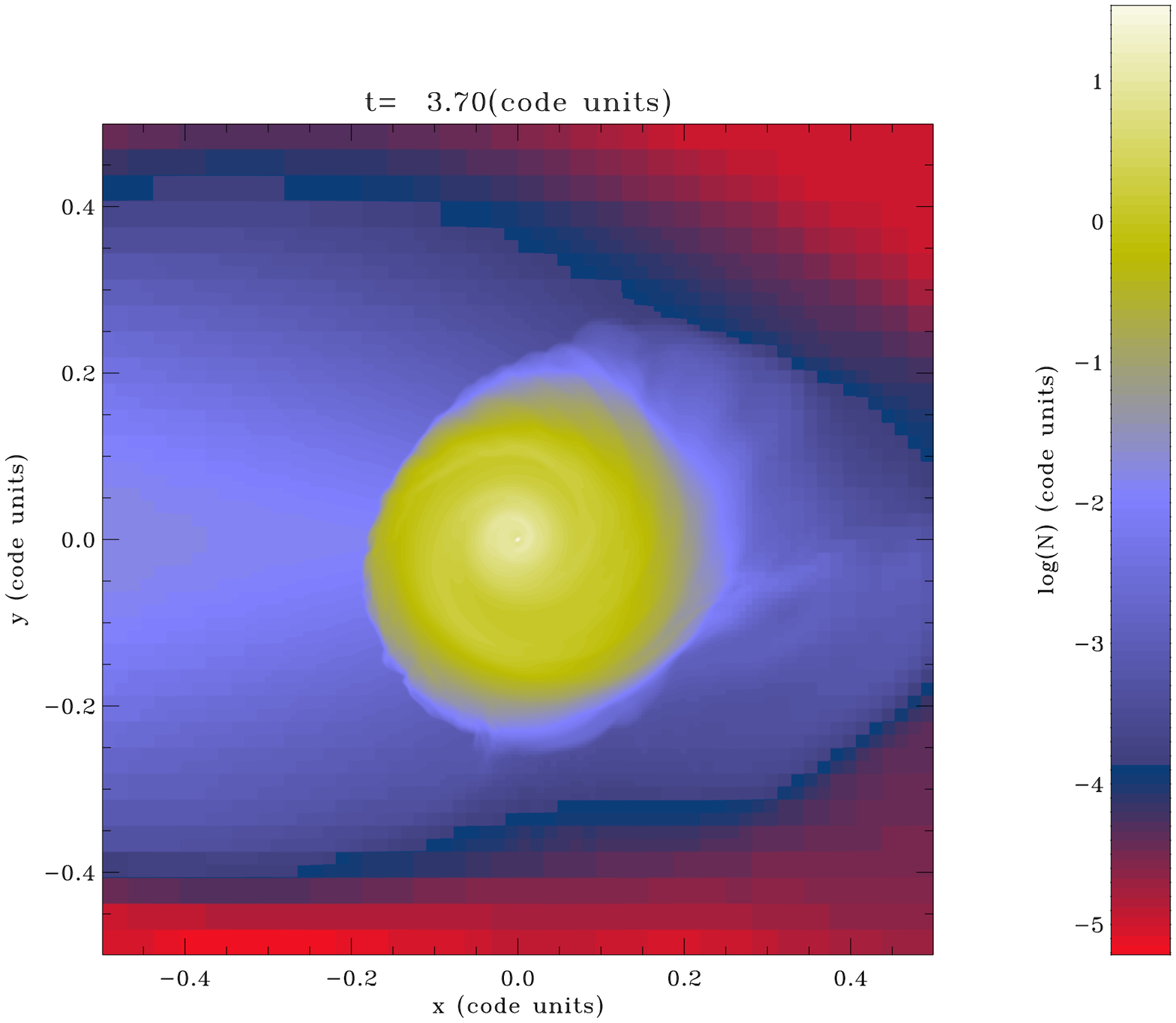}}  
\put(0,0){\includegraphics[width=8cm]{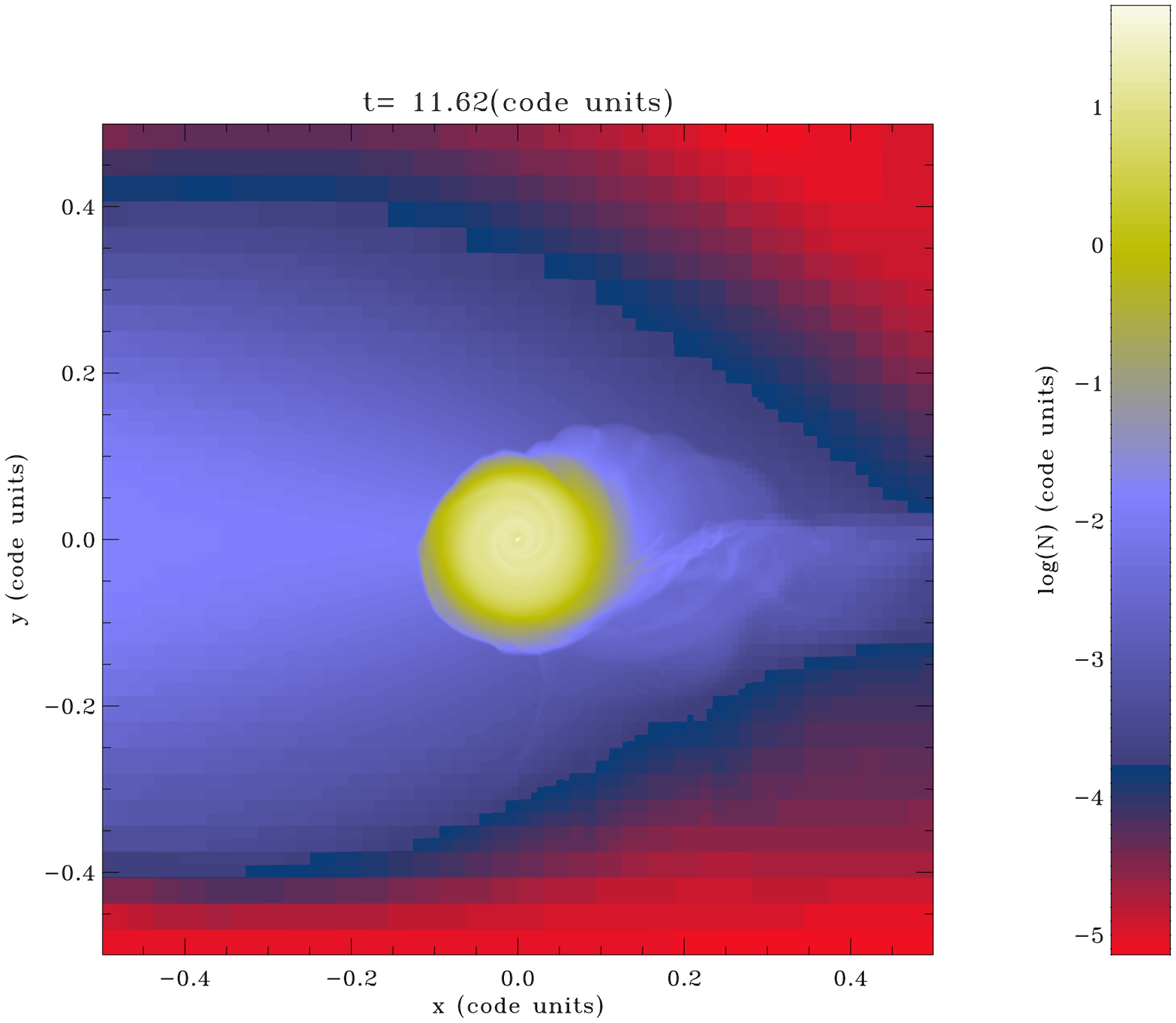}}  
\end{picture}
\caption{Simulation ASNR2 (asymmetrical accretion with $U_0=2$ and no rotation). Column density map
at 2 different timesteps. The disc quickly shrinks in size.}
\label{accret_no_rot_map}
\end{figure}

Finally since angular momentum is another quantity which is playing 
an important role and whose distribution is not well understood, Fig.~\ref{accret_no_rot_map} presents 
a calculation in which the accreted gas is injected without angular momentum. As can be seen the impact on the disc
is really drastic since its radius shrinks by almost a factor 2. This is because on one hand the centrifugal radius 
is proportional to $j^2$, $j$ being the specific angular momentum,  and on the other hand, the specific 
angular momentum decreases because the total mass increases but not the total angular momentum. 
At the time corresponding to the second snapshot ($t=11.62$), the mass in the box has increased by about 40$\%$. Therefore the 
specific angular momentum has decreased by the same amount and since $1.4^2 \simeq 2$, we get a reasonable agreement. 
The $\alpha$ values, not been displayed here for conciseness, are also much higher than for the other cases and typically reach around $0.1$.   
Interestingly, although the disc dynamics appears to be rather vigorous, there are no sign of the prominent spiral pattern 
which develops in ASR2 in the vicinity of the disc.  As discussed above we believe that this is due to the absence of rotation 
within the accreting gas. \\

These two cases clearly illustrate that the accretion rate is not the only parameter playing  a role. There is therefore 
considerable uncertainty and there may be a broad diversity of situations depending on the real distribution of 
accretion episode in star formation regions.

\section{Discussion}

\subsection{Interpretations of  observations}
\label{discus}

\setlength{\unitlength}{1cm}
\begin{figure} 
\begin{picture} (0,14)
\put(0,0){\includegraphics[width=8cm]{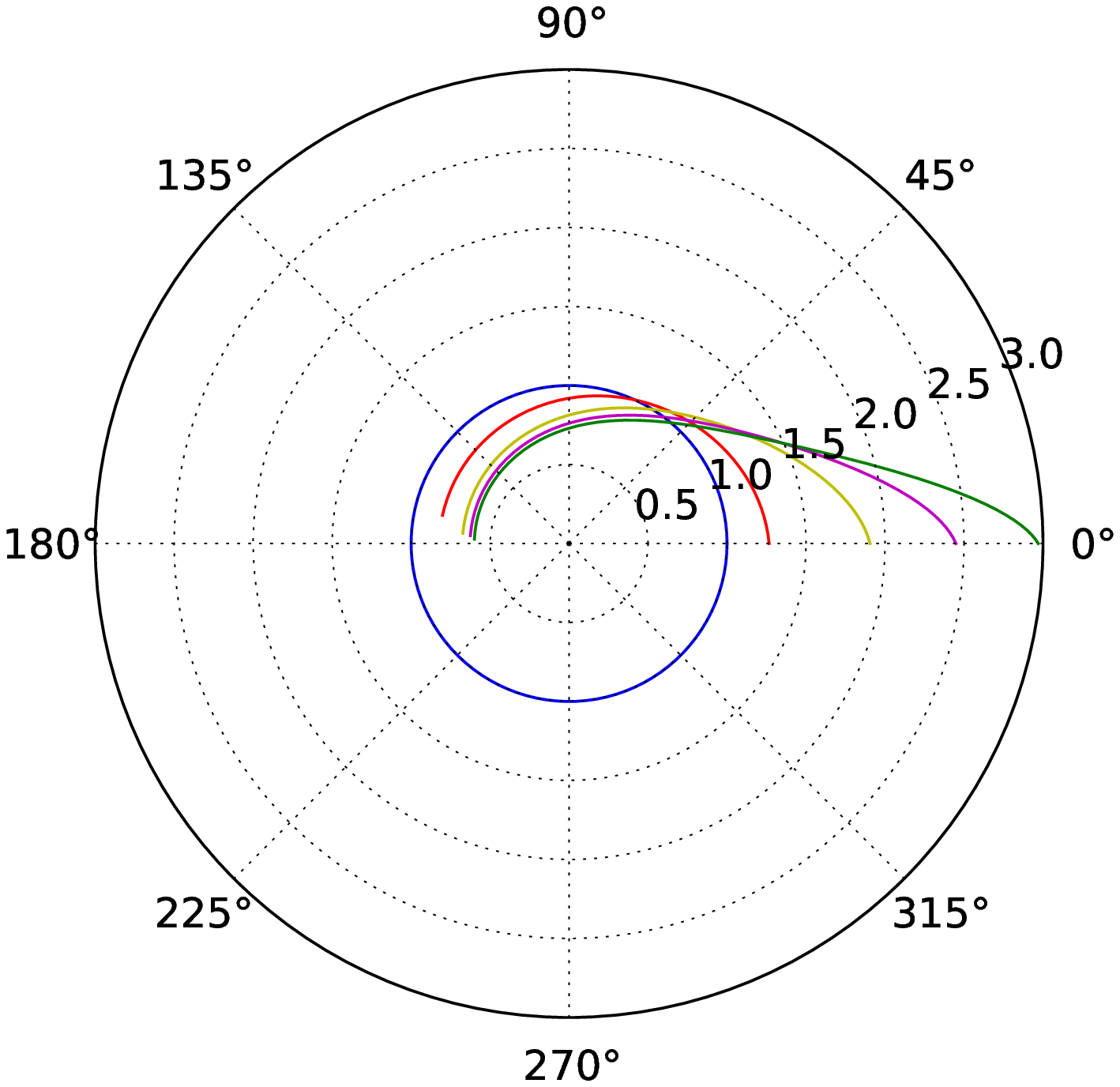}}  
\put(0,7){\includegraphics[width=8cm]{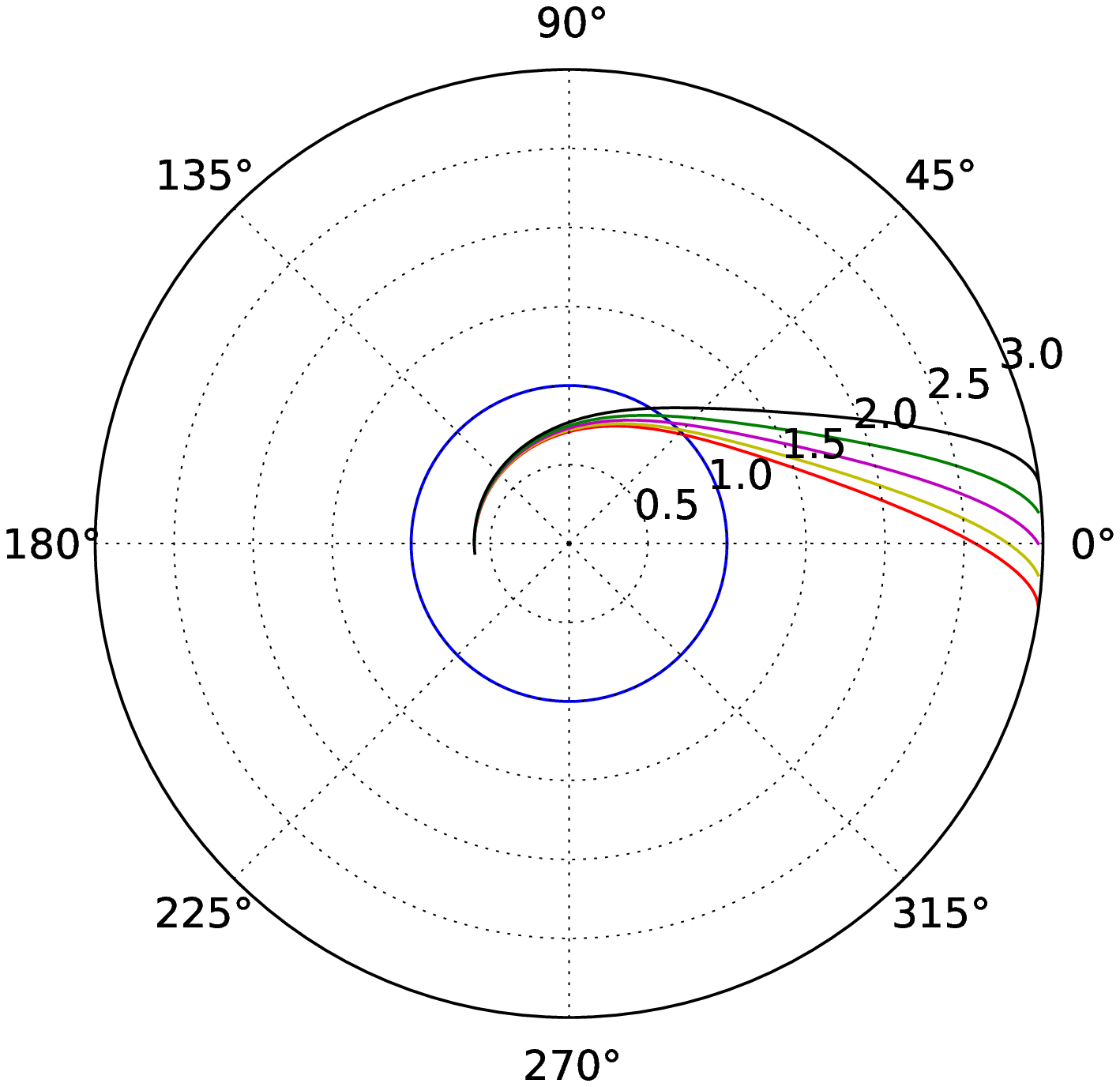}}  
\end{picture}
\caption{Elliptical orbits for various particles with same energy but different 
trajectories mimicking the accretion studied in this paper 
(top panel) and particles with different
energies (specified by $\eta=r_0/r_d$) but same 
angular momentum. As can be seen the orbits tend to approach or even cross each other which would result in a shock and a spiral arm.}
\label{stream}
\end{figure}

Many observations of evolved transition discs at various wavelengths from radio to optical, have been carried out 
in recent years \citep[e.g.][]{hashimoto+2011,garuti+2013,boccaletti+2013,christiaens+2014,avenhaus+2014,wagner+2015,benisty+2015}.
They consistently report the presence of density structures such as prominent spiral arms located within or {\it in the vicinity} 
of the disc. The origin of these structures remains debated. They have often been attributed to  the presence of a massive planet 
which indeed can excite spiral patterns \citep[e.g.][]{rafikov+2002,baruteau+2014}. Self-gravity is another mechanism 
which has been proposed to explain these waves but it is not quite clear that all the discs where spiral patterns are 
observed are massive enough \citep[see discussion in ][]{hall+2016}. 

While these explanations are certainly possible, the  simulations performed in this work suggest 
that accretion is another viable mechanism. Indeed, visual inspection and comparison between 
Fig.~\ref{col_map_2} and figures such as Fig.~1 of \citet{hashimoto+2011} or Fig.~1 of \citet{benisty+2015} 
show striking resemblance with clear spiral patterns that tend to open up as they get more distant from the 
disc. In the simulation, at least, these spiral features are clearly a consequence of the accretion of rotating gas 
that is approaching centrifugal equilibrium (let us recall that these spiral arms are absent in the non-rotating accretion case
shown in Fig.~\ref{accret_no_rot_map}). To understand this more quantitatively let us simply follow a rotating fluid 
particle that is infalling toward a star. Let us assume that its specific angular momentum, $j$ is conserved and let us ignore all 
forces but gravity and centrifugal acceleration. 
Angular momentum conservation and energy conservation lead to
\begin{eqnarray}
\nonumber
v_\phi = {j \over r}, \\
{1 \over 2} (v_r^2 + v_\phi^2) - {G M_* \over r} = - {G M_* \over r_0}.
\label{conserv} 
\end{eqnarray}
Combining them, we can express 
\begin{eqnarray}
{v_r \over v_\phi} = \left( 2 \left( {r \over r_d} - {r^2 \over r_0 r_d} \right) -1 \right) ^{1/2} = \tan \theta,
\label{tantheta}
\end{eqnarray}
where $r_d=j^2/G M_*$ is the centrifugal radius and where 
$\theta$ is the angle between the stream line and the azimuthal direction.
Let us stress that the trajectory are simply Keplerian orbits of various eccentricity.
Figure~\ref{stream} shows five orbits corresponding to the same angular momentum and energy 
but different trajectories mimicking the accretion used in the simulations presented in this paper
(upper panel)
and five orbits with different energies 
(leading to different values of $r_0 = \eta r_d$ where $r_d$ is the Keplerian radius and corresponds to
the blue circle)  which would result from energy dissipation 
in the accretion flow (lower panels). 
The five values of $\eta$ are 1, 2.1, 2.6, 3.1 and 3.6. 
Interestingly the orbits get very close to each others and even cross, which would unavoidably 
result in a shock and a spiral structure. It is tempting to attribute the formation of the 
prominent spiral structures seen in Fig.~\ref{col_map_2} to this mechanism. These spirals 
could therefore be simply nearly kinematic waves \citep{toomre77}.

From Eq.~(\ref{tantheta}), we see that typically the stream line makes an angle $\theta \simeq 45^\circ $ just before 
it enters the disc. This angles increases at larger radii and for example is about 60$^\circ$
at $r=2 r_d$. Note that if energy is not conserved, which is likely the case close to the disc because of the shock
then the  angle $\theta$ is accordingly reduced. Moreover let us stress that the disc radius is not so well determined
in particular because the incoming particles are shocking on its surface (compare Figs.~\ref{no_accret_map} and \ref{col_map_2}),
therefore the spiral arm may be able to penetrate ``inside'' the disc as is the case in Fig.~\ref{stream} and in 
simulations ASR2.

If this interpretation of the origin of some of the observed spiral pattern is confirmed, 
this would constitute a signature  of ongoing accretion in transitional discs.

\section{Conclusion}

We have investigated the influence that infall may have onto low mass circumstellar 
protoplanetary discs by performing 3D hydrodynamical numerical simulations
ignoring self-gravity. The underlying idea, here, is to concentrate only 
on hydrodynamics in order to disentangle the effects of the different 
physical processes. 

Our results confirm the 2D simulation results presented in \citet{lesur+2015}: infall 
onto the disc has an influence, even at small radii.  For an infall 
rate inside the computational box of about $4 \times 10^{-7}$ M$_\odot$ yr$^{-1}$ (which corresponds to an effective accretion rate onto the disc
about 3-4 times smaller), we obtained $\alpha$ values, 
which inside the disc (that is at radii about 5 times smaller than the disc radius), 
are as high as $5 \times 10^{-3}$ when the disk has a mass on the order of $10^{-2}$ M$\odot$ and a typical thickness $H/R \simeq 0.15$. At the disc outer edge, values of $\alpha$ up to $10^{-1}$ are obtained.
These values decrease with the infall rate and for $10^{-7}$ M$_\odot$ yr$^{-1}$ (corresponding to an infall onto the disc about 4 times 
smaller), we 
obtain $\alpha$ values typically 3-5 times smaller that is to say equal to about $10^{-3}$.

As these $\alpha$ values are a few times larger than the ones that have been measured by 
\citet{lesur+2015}, we have estimated the contribution to $\alpha$ of the vertically averaged flows.
The associated stress, $\alpha_{2D}$, was compared to the quantity 
$\alpha_z = \alpha - \alpha_{2D}$, which quantifies the transport associated with 
3D structures. We found that $\alpha_{2D}$ and $\alpha_z$ have similar values in the inner part of the disc.
 This suggests that 3D dynamical 
processes, in particular the fact that the accreted gas can fall into the disc from above and below rather than 
from the equatorial plane only, is playing a role and produces larger fluctuations.
For the conditions that we have been exploring, we found that typically about 10$\%$ of the mass injected 
in the computational box
 is  accreted deep inside the disc (at a radius smaller than half its initial radius).
Since the mass which falls effectively into the disc is 3-4 smaller than the  mass injected into the box, 
 the mass that is being accreted in the inner part of the disc is about half the mass that effectively falls
 onto it, though defining the disc boundary accurately appears to be a difficult task. 

Finally, we have also explored the influence of the symmetry and the angular momentum 
of the accreted gas, finding that both have a significant influence on the effects
that infall has onto the disc. More symmetrical flows have a more limited influence while 
slowly rotating material may influence the disc dramatically since it reduces the specific
angular momentum.  
This clearly indicates that  precisely quantifying the impact that infall may have on 
discs requires a detailed knowledge of the star formation process and the nearby 
environment of the star-disc system.

The accretion of rotating gas, naturally produces prominent spiral patterns both inside and outside the disc.
The latter seems to be a natural consequence of spiralling collapsing gas dynamics and largely triggers the 
spiral pattern within the disc. This pattern presents striking resemblance with recent observations of transitional discs, 
which may indicate that, at least some of the spirals observed in these objects are due to infalling motions rather than 
due to the presence of planets. \\ \\

\emph{Acknowledgments}
This work was granted access to HPC resources of CINES under the 
allocation  x2014047023 made by GENCI (Grand Equipement National de Calcul Intensif).
This research has received funding from the European Research Council under the European
 Community's Seventh Framework Programme (FP7/2007-2013 Grant Agreement no. 306483).

\bibliographystyle{aa}
\bibliography{biblio}

\end{document}